\documentclass[a4paper,11pt,oneside]{article} 
\usepackage{jheppub_noruler}
\makeatletter
\def\@fpheader{}
\makeatother

\notoc
\preprint{RBI-ThPhys-2020-06}
  
\dedicated{\centerline{Talk presented at
  the International Workshop on Future Linear Colliders (LCWS2019)}
  \centerline{Sendai, Japan, 28 October-1 November, 2019.}
  \centerline{C19-10-28} }

\renewcommand\afterEmailSpace{ \\[1cm]}
\renewcommand\afterAbstractSpace{ \\[2cm] }

\usepackage{graphicx}  
\graphicspath{{./plots/}}

\usepackage{xspace}
\newcommand{\whizard}{\textsc{Whizard}\xspace}
\newcommand{\delphes}{\textsc{Delphes}\xspace}

\title{Searching Inert Scalars at Future  e$^+$e$^-$ Colliders}

\author[a\star]{Aleksander~Filip~\.Zarnecki,}
\author[a]{Jan Kalinowski,} 
\author[a]{Jan Klamka,}
\author[a]{Pawel Sopicki,}
\author[b]{\\Wojciech Kotlarski,}
\author[c]{Tania Robens,}
\author[d]{Dorota Sokolowska}
\affiliation[a]{Faculty of Physics, University of Warsaw, Warsaw, Poland}
\affiliation[b] {Institut f\"ur Kern- und Teilchenphysik, TU Dresden, Dresden, Germany}
\affiliation[c]{Theoretical Physics Division, Rudjer Boskovic Institute, Zagreb, Croatia}
\affiliation[d]{International Institute of Physics, Universidade Federal do Rio Grande do Norte, Brazil}
\affiliation[\star]{Speaker}
\emailAdd{filip.zarnecki@fuw.edu.pl}

\abstract{
The Inert Doublet Model (IDM) is one of the simplest extensions of the
Standard Model (SM), providing a dark matter candidate.
It is a two Higgs doublet model with a discrete $Z_2$ symmetry, that
prevents the scalars of the second doublet (inert scalars) from coupling to the
SM fermions and makes the lightest of them stable.
We study a large number of IDM scenarios, which are consistent
with current constraints on direct detection and relic density of dark
matter, as well as with all collider and low-energy limits.
We propose a set of benchmark points with different kinematic
features, that promise detectable signals at future $e^+e^-$
colliders.
Two inert scalar pair-production processes are considered, $e^+e^- \to
A~H $ and $e^+e^- \to H^+H^-$, followed by decays of $A$ and $H^\pm$
into final states which always include the lightest and stable neutral
scalar dark matter candidate $H$.
Significance of the expected observations is studied for different
benchmark models and different running scenarios, for centre-of-mass
energies from 250 GeV up to 3\,TeV.
For low mass scenarios, high significance can be obtained for the
signal signatures with two muons or an electron and a muon in the
final state.
For high mass scenarios, which are only accessible at high energy
stages of CLIC, the significance is too low for the leptonic signature
and the semi-leptonic final state has to be used as the discovery
channel.
Results presented for this channel are based on the
fast simulation of the CLIC detector response with the \delphes
package.
}

\begin{document} 

\maketitle 



\section{Inert Doublet Model}

A number of astrophysical observations based on gravitational interactions point to
the existence of dark matter (DM) in the Universe, which can not be described with
the Standard Model.
One of the simplest extensions of the Standard Model which can
provide a dark matter candidate is the Inert Doublet
Model~\cite{Deshpande:1977rw,Cao:2007rm,Barbieri:2006dq}. 
In this model, the scalar sector is extended by a so-called inert or
dark doublet $\Phi_D$ (the only field odd under $Z_2$ symmetry) in
addition to the SM Higgs doublet $\Phi_S$. This results in five
physical states after electroweak symmetry breaking: the SM Higgs
boson $h$ and four dark scalars:  two neutral, $H$ and $A$, and two
charged,  $H^\pm$. 
A discrete $Z_2$ symmetry prohibits the inert scalars from interacting
with the SM fermions through Yukawa-type interactions and makes the
lightest neutral scalar, chosen to be $H$ in this work, a good dark
matter candidate.

Two sets of benchmark points (BPs) in agreement with all
theoretical and experimental constraints were proposed
in~\cite{Kalinowski:2018ylg}, covering different possible signatures
at $e^+e^-$ colliders, with masses of IDM particles extending up to
1 TeV. 
Prospects for the discovery of IDM scalars at CLIC running at
380\,GeV, 1.5\,TeV and 3.5\,TeV were then described in detail
in~\cite{Kalinowski:2018kdn} and summarized in \cite{deBlas:2018mhx},
focusing on leptonic final states.  
In this contribution we update these results
and extend them to ILC running at 250\,GeV and 500\,GeV. 
We also include newe results based on the semi-leptonic channel analysis,
for CLIC running at  1.5\,TeV and 3.5\,TeV, which supersede results
presented in \cite{Sokolowska:2019xhe}. 

\section{Benchmark scenarios}

\begin{figure}[b]
\includegraphics[width=0.49\textwidth]{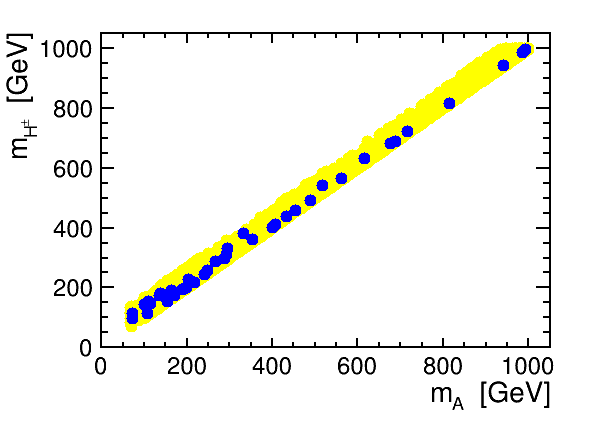}
\includegraphics[width=0.49\textwidth]{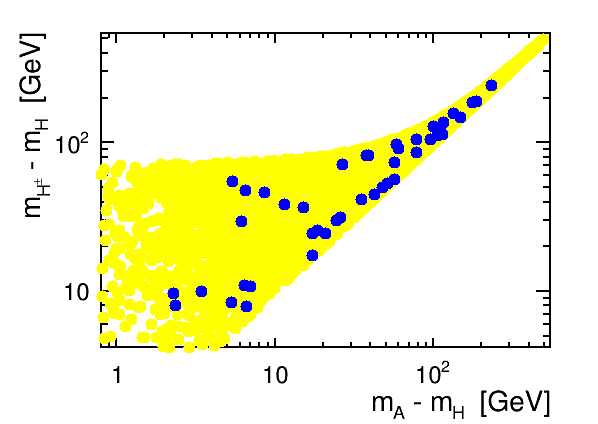}
\caption{ Distribution of benchmark candidate points (yellow) in the
  (m$_{A}$;m$_{H^\pm}$) plane (left) and in the
  (m$_{A} -\,$m$_{H}$;m$_{H^\pm} -\,$m$_{H}$) plane (right),
after all constraints are taken into account, as well as selected
benchmark points (blue) in the same planes~\cite{Kalinowski:2018ylg}.
}\label{fig:mass}
\end{figure}

Distributions of the scalar masses for the IDM benchmark scenarios
considered in~\cite{Kalinowski:2018ylg} are shown in Fig.~\ref{fig:mass}.
For the considered benchmark scenarios $H$ is the lightest, stable neutral
scalar, which can be much lighter than the other two, $A$ and
$H^\pm$.
On the other hand the mass splitting between $A$ and
$H^\pm$ is limited by existing constraints to about 70\,GeV.

\begin{figure}[t]
\includegraphics[width=0.49\textwidth]{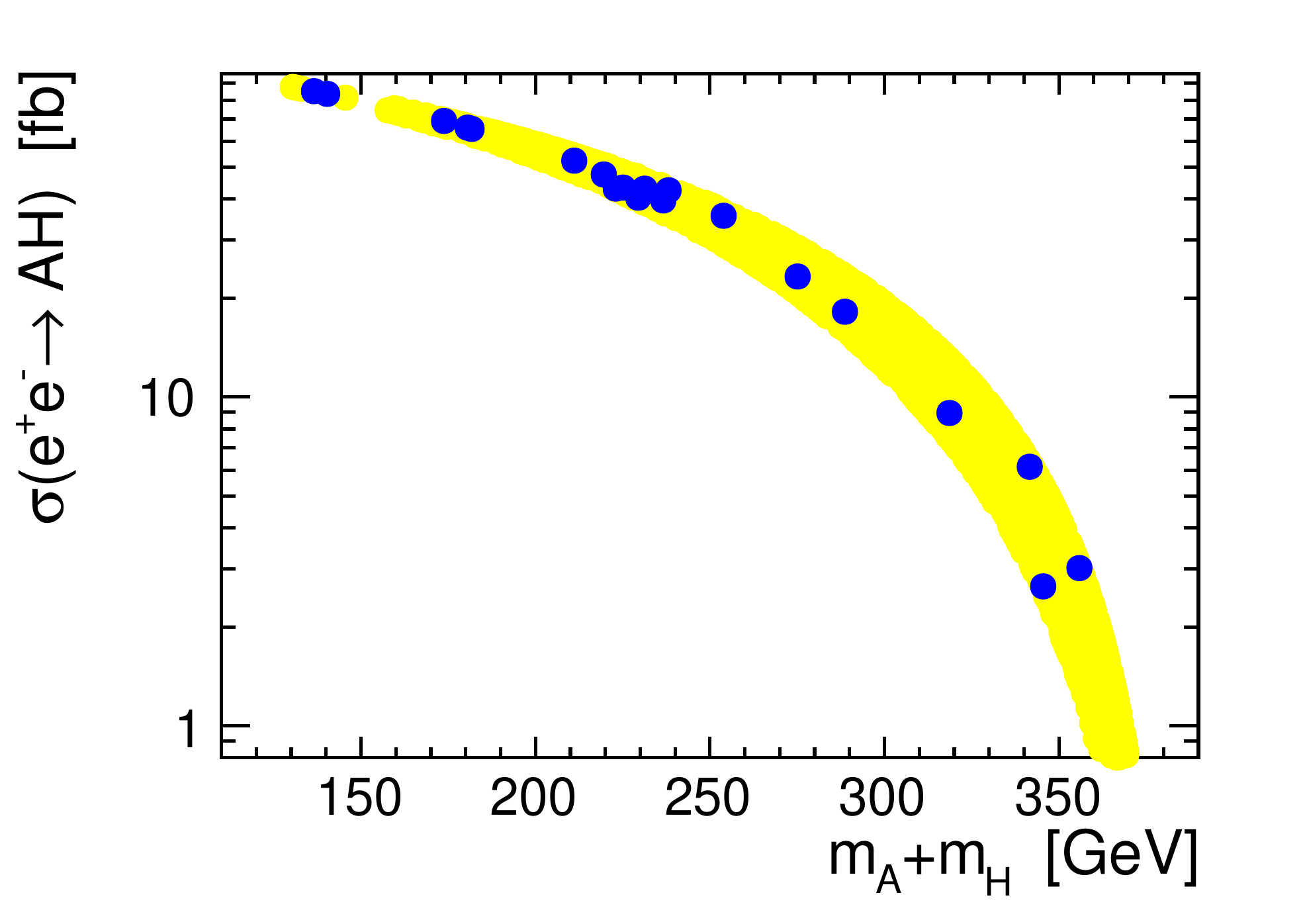}
\includegraphics[width=0.49\textwidth]{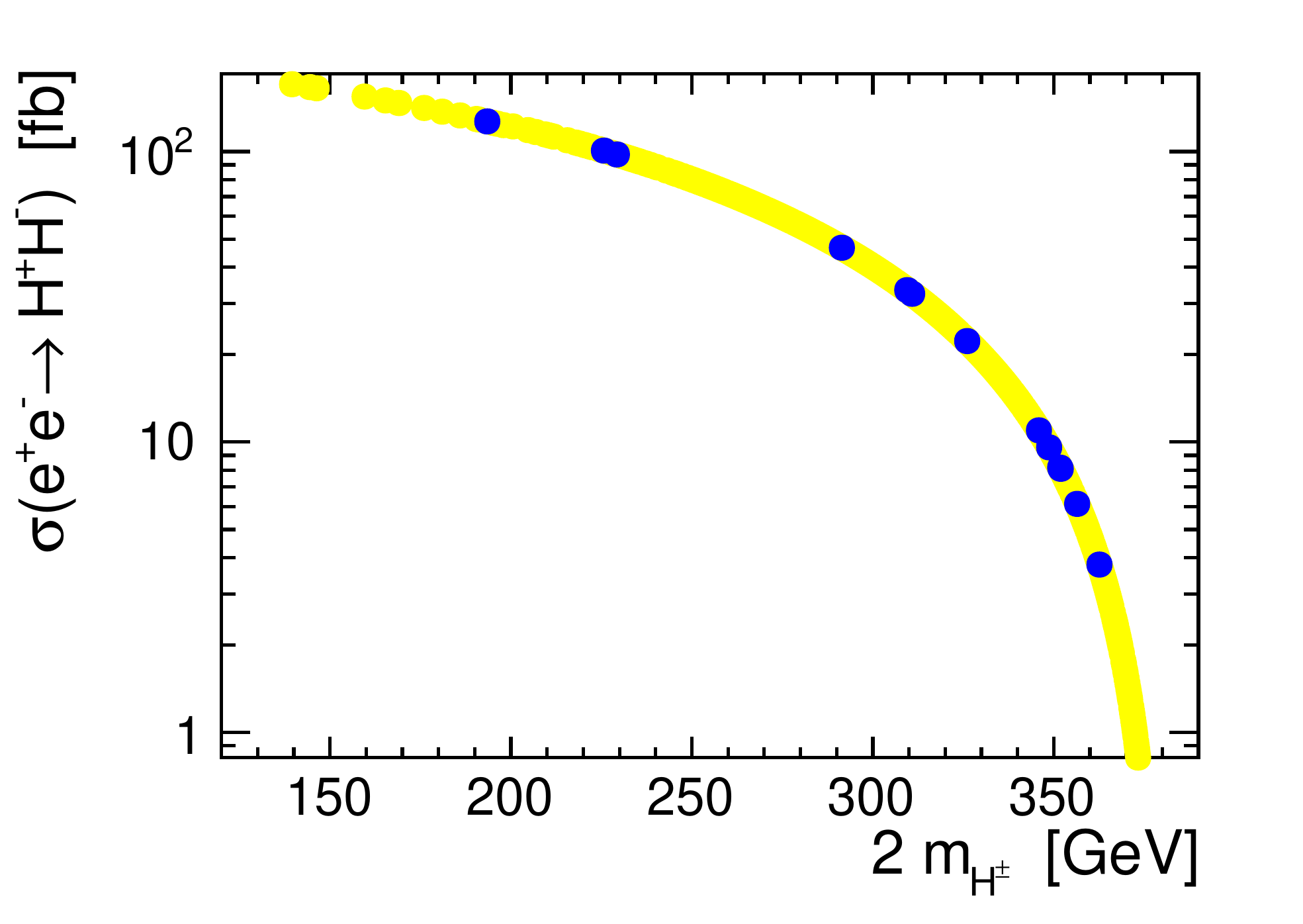}\\[-1.7cm]
                {\hspace*{0.1\textwidth}\scriptsize $\sqrt{s}=$380 GeV}\\[-0.53cm]
                {\scriptsize \hspace*{0.59\textwidth} $\sqrt{s}=$380 GeV} \\[0.95cm]
\includegraphics[width=0.49\textwidth]{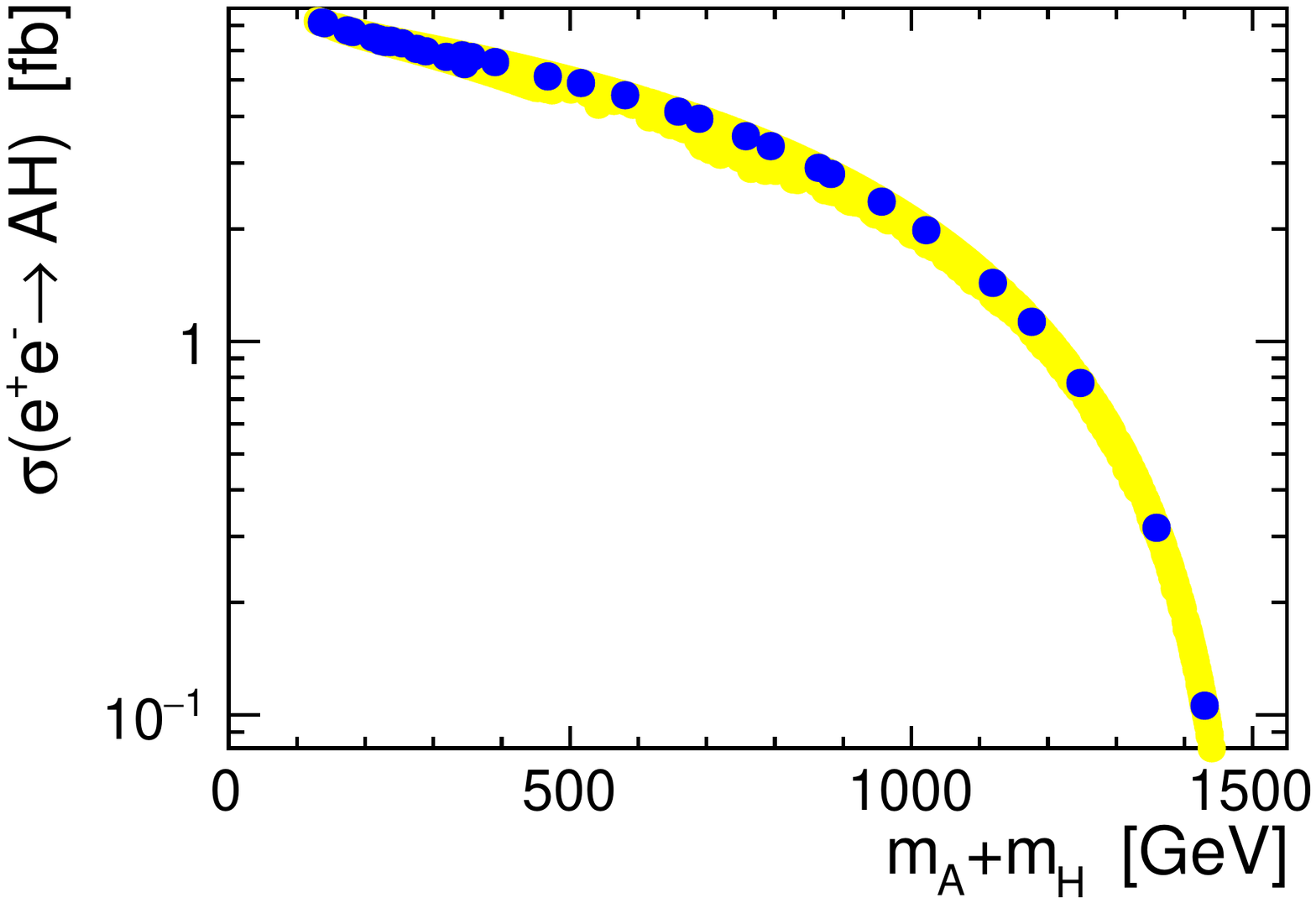}
\includegraphics[width=0.49\textwidth]{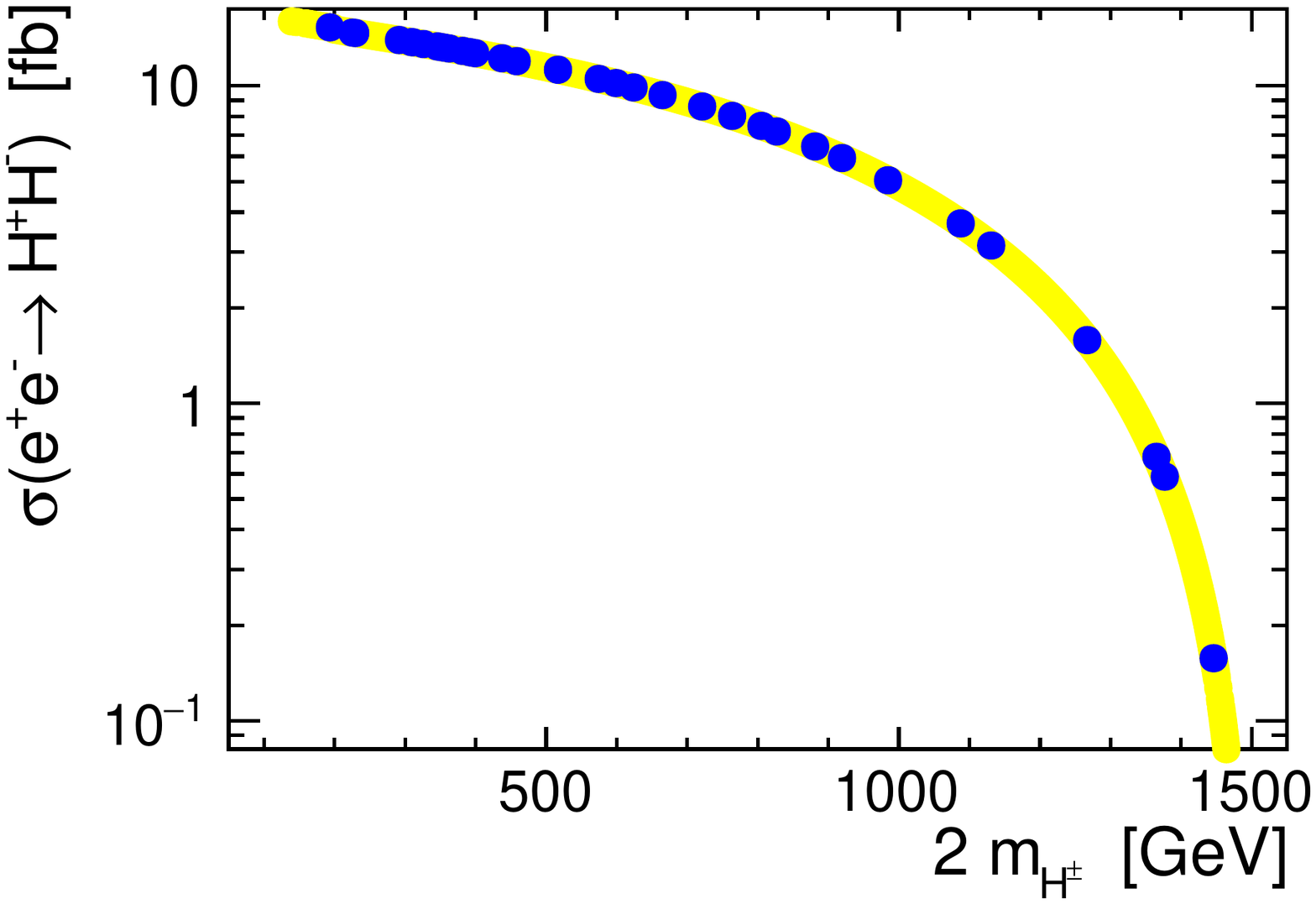}\\[-1.7cm]
                {\hspace*{0.1\textwidth}\scriptsize $\sqrt{s}=$1.5 TeV}\\[-0.53cm]
                {\scriptsize \hspace*{0.59\textwidth} $\sqrt{s}=$1.5 TeV} \\[0.5cm]
\caption{ Leading-order cross sections for neutral (left) and charged (right) inert
  scalar production,  $ e^+e^- \to A~H$ and $e^+e^-\to H^+H^-$, for collision energy
  of 380\,GeV (upper plots) and 1.5\,TeV (lower plots).  The yellow band represents
  all scenarios selected in the model scan~\cite{Kalinowski:2018ylg} while the 
blue dots represent the selected benchmark scenarios. Beam energy spectra are not
included.}\label{fig:cros}
\end{figure}

The following tree-level production processes of inert scalars
at $e^+ e^-$ collisions are considered: neutral scalar
pair-production, $ e^+e^- \to A~H$, and charged scalar
pair-production, $e^+e^-\to H^+H^-$.
The leading-order cross sections for these processes are presented in
Fig.~\ref{fig:cros} for collision energies of 380\,GeV and 1.5\,TeV. 
As the couplings of inert scalars to SM bosons are determined by SM parameters,
production cross sections are determined by the scalar masses and
depend very weakly on additional model parameters.
Far from the kinematic threshold, the production cross section,
dominated by the $s$-channel $Z$-boson exchange, decreases as $1/s$
with the collision energy. 

In the scenarios considered in this paper the produced dark scalar $A$
decays to a (real or virtual) $Z$ boson and the (lighter)
neutral scalar $H$, $A \rightarrow Z^{(\star)}H$, while the produced
charged boson $H^\pm$ decays predominantly to a (real or virtual) $W^\pm$ boson
and the neutral scalar $H$, $H^+ \rightarrow {W^\pm}^{(\star)}H$,
where the DM candidate $H$ escapes detection.
The mono-$Z$ signature of the neutral scalar pair-production can be
considered in the leptonic or hadronic $Z$-boson decay channel.
For the charged scalar pair production, resulting in two $W$ bosons in
the final state, leptonic, semi-leptonic and hadronic final states are possible.

\section{Leptonic channel analysis}
\label{sec:leptonic}

\begin{figure}[t]
\centerline{\includegraphics[width=0.75\textwidth]{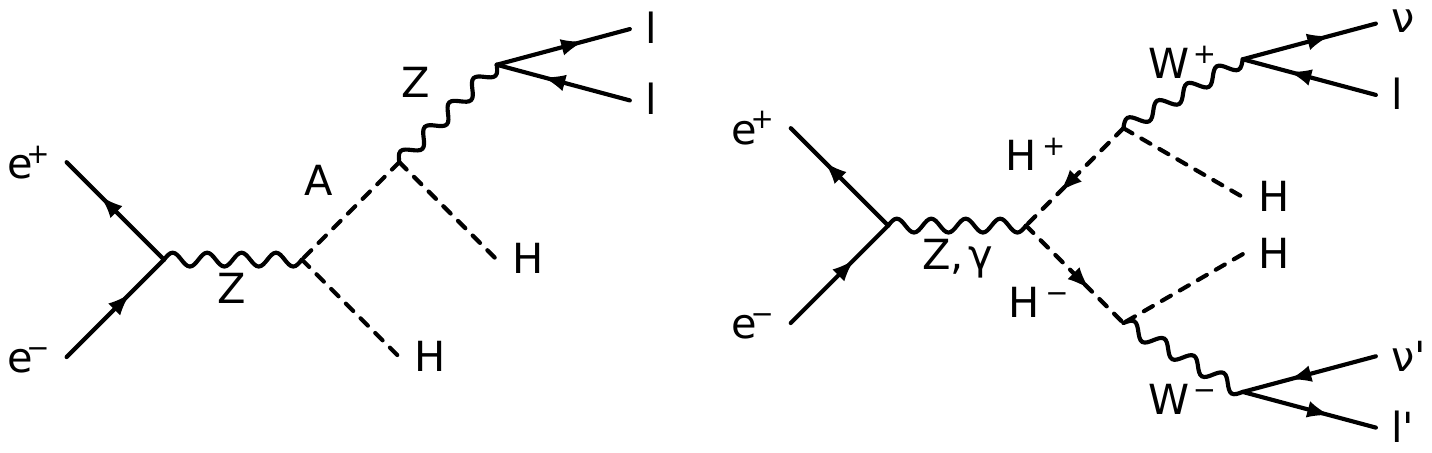}}
\caption{Signal Feynman diagrams for the considered production and
  decay process for:
(left) neutral scalar production, $e^+e^- \to H A \to H H l l$,
and
(right) charged scalar production, $e^+e^- \to H^+ H^- \to H H l l' \nu \nu'$.
}\label{fig:diag}
\end{figure}

Isolated leptons (electrons and muons) can be identified and
reconstructed with very high efficiency and purity, and the signatures
based solely on lepton measurements are usually considered ``golden
channels'', if the expected statistics of signal events is high
enough.
For purely leptonic final state, the detector acceptance cuts can be
applied on the generator level and other detector effects are expected
to have marginal impact on the outcome of the analysis.
Therefore, in~\cite{Kalinowski:2018kdn} we focused  on
leptonic decays of  $Z$ and $W^\pm$,
leading to a signature of leptons and missing transverse energy.
We considered the $\mu^+\mu^-$ final state as a
signature of the neutral scalar pair-production, while
the different flavour lepton pairs, $\mu^+ e^-$ and $e^+ \mu^-$, were
considered as a signature for production of charged inert scalars, see
Fig.~\ref{fig:diag}. 

\begin{figure}[tb]
\includegraphics[width=0.49\textwidth]{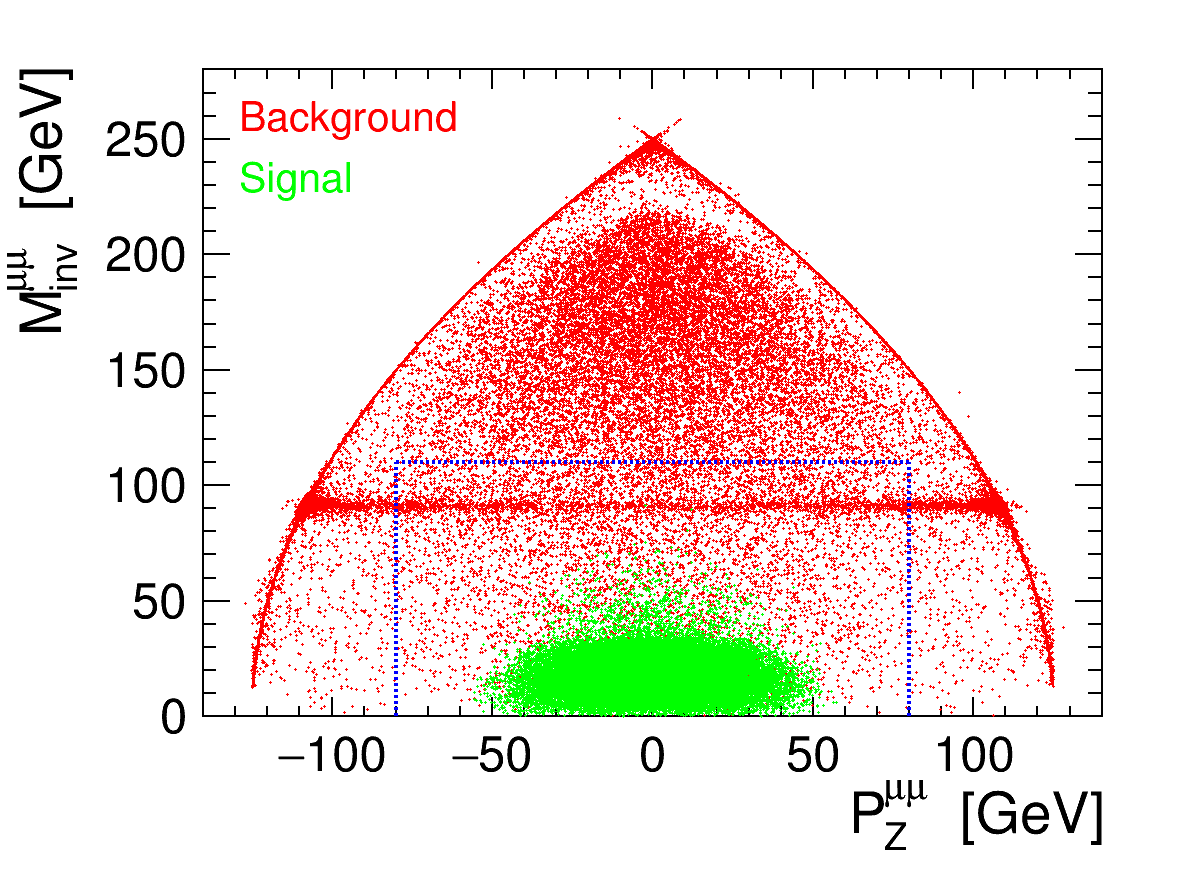}
\includegraphics[width=0.49\textwidth]{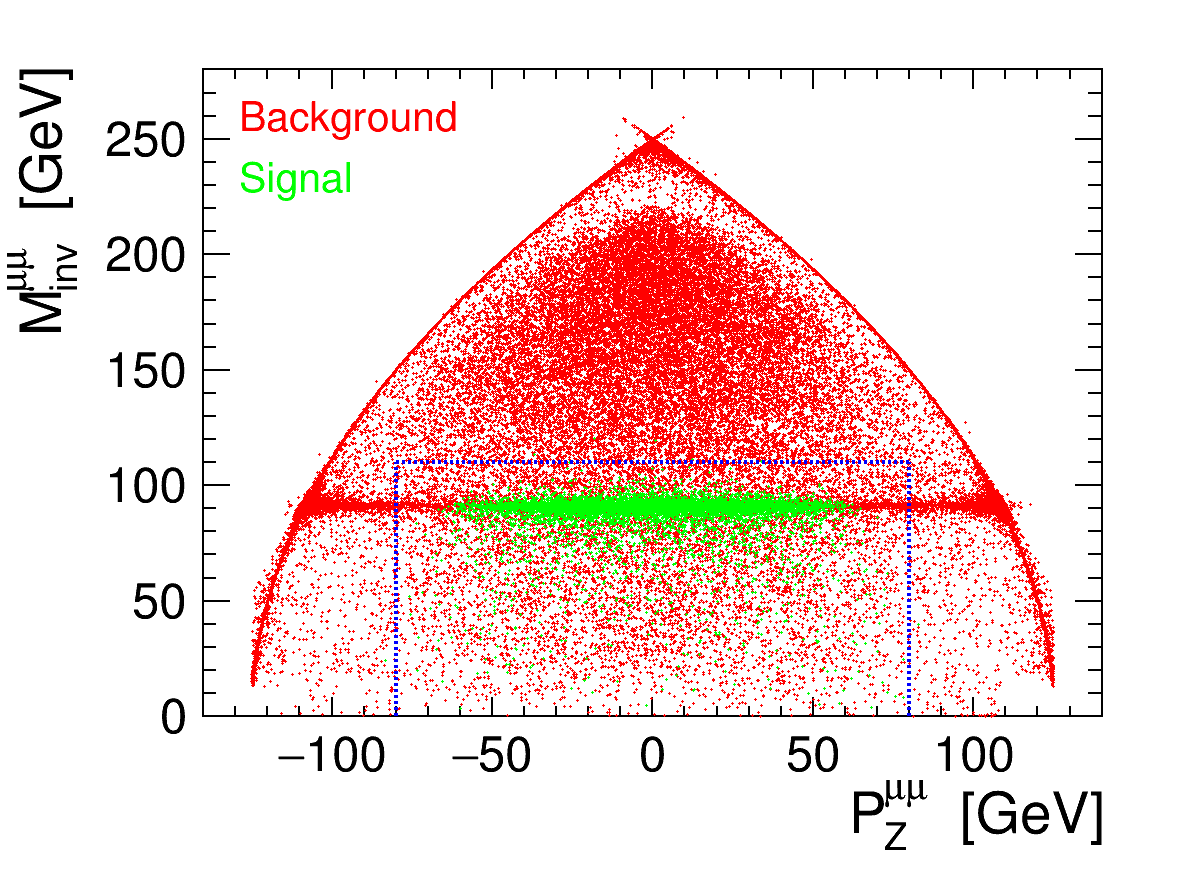}
\caption{
  Distribution of the lepton pair invariant mass, M$_{\mu\mu}$,
  as a function of the lepton pair longitudinal momentum,
  P$_\textrm{z}^{\mu\mu}$, for IDM signal (green points) and Standard Model
  background (red points). Signal events were simulated for BP1
  scenario (left) and BP9 scenario (right), for centre-of-mass energy
  of 250\,GeV. The blue box indicates the cut used to remove the
  dominant background from $e^+e^- \to \mu^+\mu^- (\gamma)$ process.
}\label{fig:presel}
\end{figure}

\begin{figure}[tb]
\includegraphics[width=0.49\textwidth]{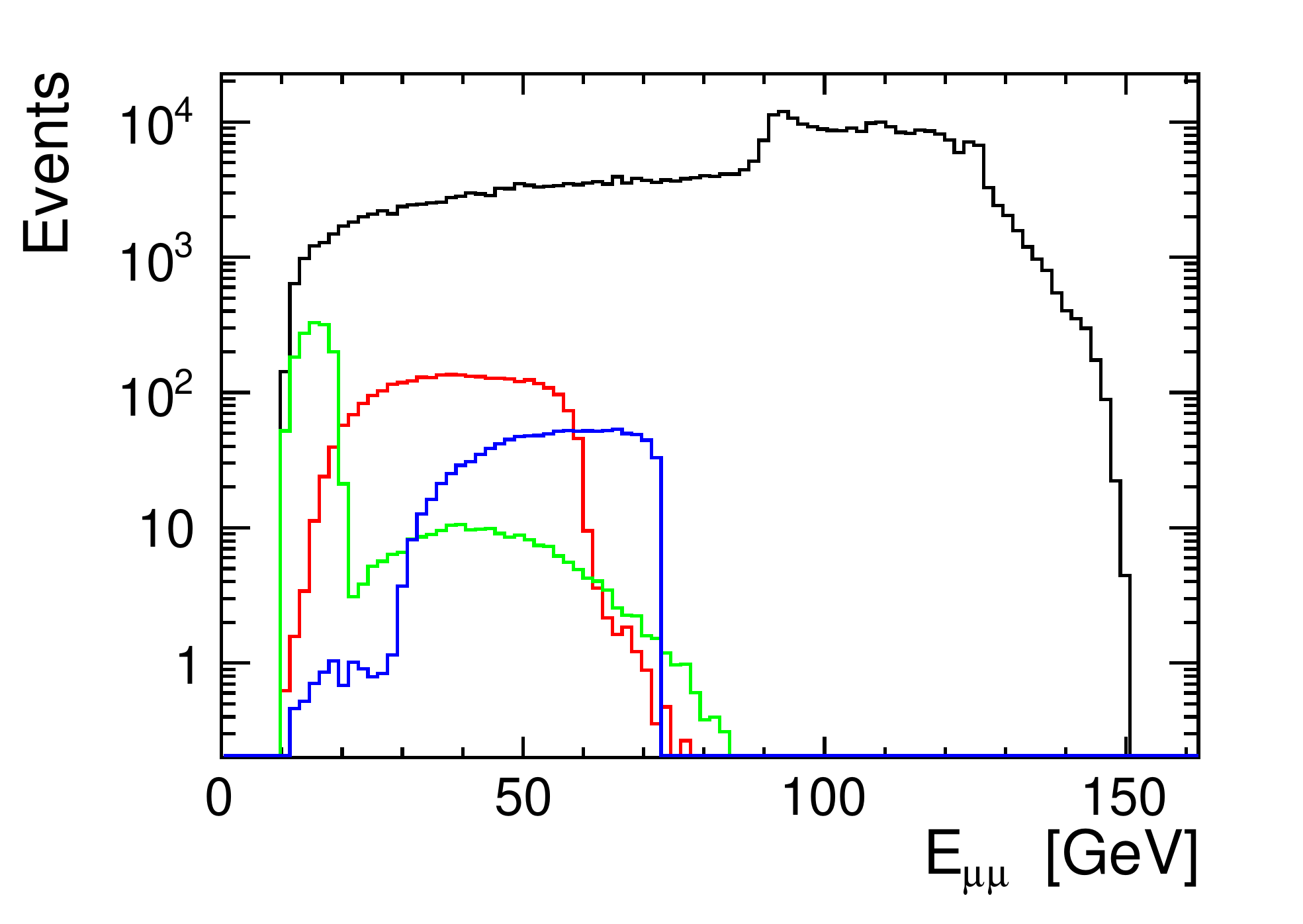}
\includegraphics[width=0.49\textwidth]{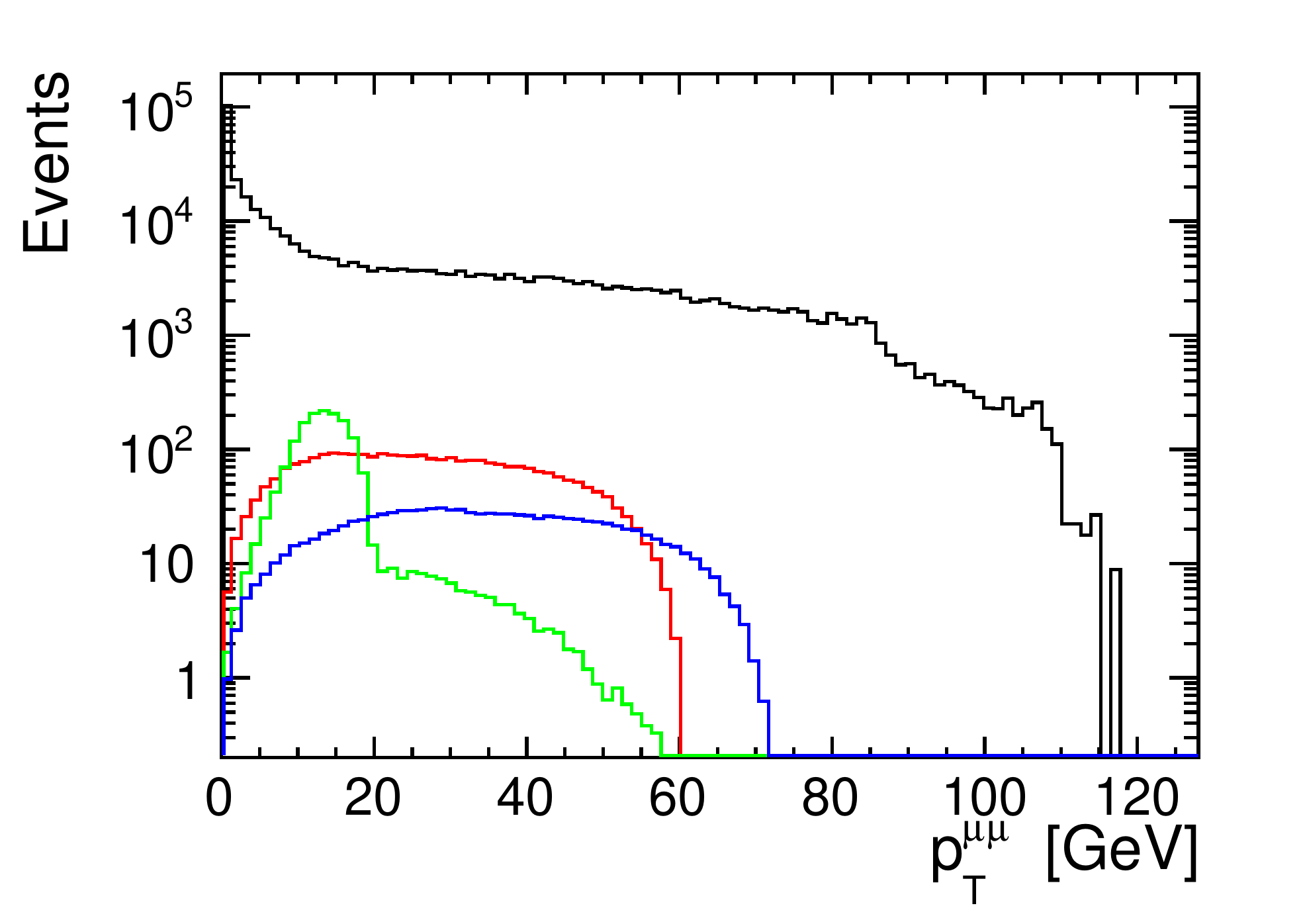}
\caption{
  Distributions of the kinematic variables describing the
  leptonic final state in $AH$ analysis: lepton pair
energy, E$_{\mu\mu}$ and total transverse momentum, p$^{\mu\mu}_\textrm{T}$.
Expected distributions for representative benchmarks BP1 (red
histogram), BP2 (green) and BP7 (blue) are compared with expected
background (black histogram) simulated for 1\,ab$^{-1}$ collected at 250\,GeV.
}\label{fig:dist}
\end{figure}

\begin{figure}[tb]
\includegraphics[width=0.49\textwidth]{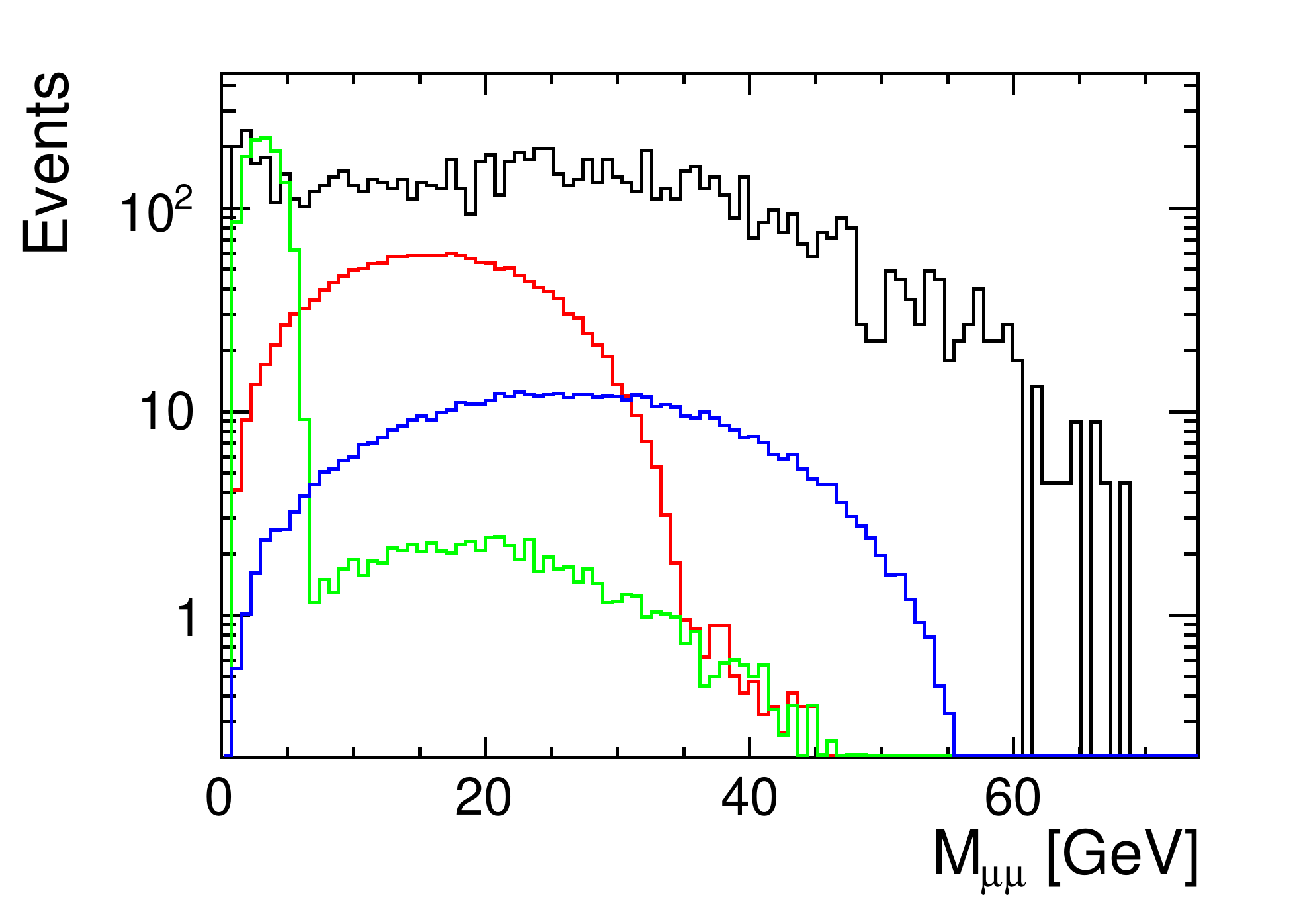}
\includegraphics[width=0.49\textwidth]{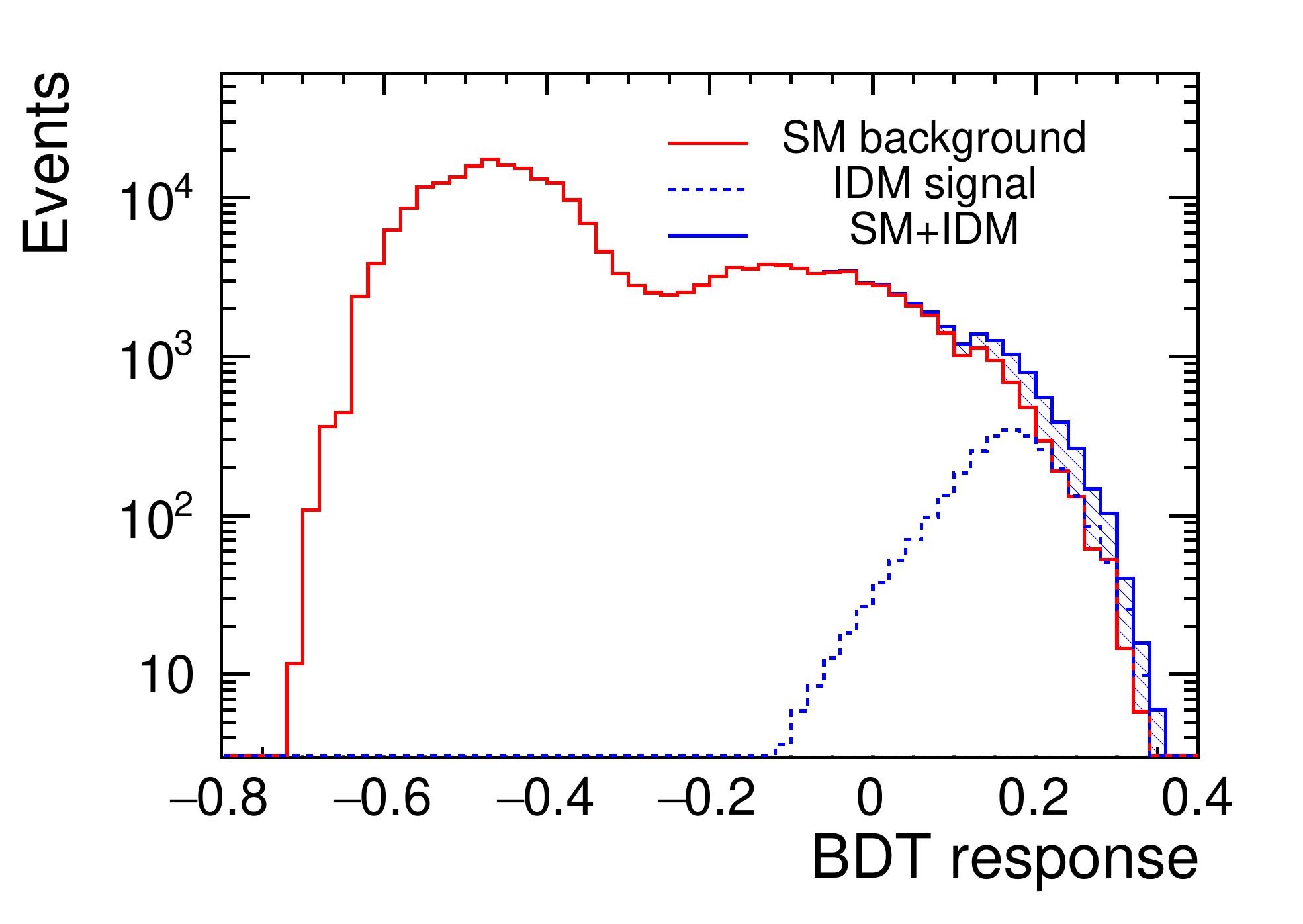}
\caption{
  Left: distribution of the lepton pair invariant mass, M$_{\mu\mu}$,
  for BP1 (red histogram), BP2 (green) and BP7 (blue) signal
  scenarios, compared with the expected Standard Model background (black
  histogram), after event  selection cuts (see text for details).
  Right: response distributions of the BDT classifiers used for the
  selection of $AH$ production events, for BP1.
  Samples are normalised to 1\,ab$^{-1}$ collected at 250\,GeV.
}\label{fig:bdt}
\end{figure}

Signal and background samples were generated with \whizard
2.2.8~\cite{Kilian:2007gr,Ohl:2000hq}. Generator level cuts reflecting detector
acceptance for leptons and ISR photons were applied.
For the neutral inert scalar pair production, $e^+ e^- \to AH$,
the invariant mass of the lepton pair from (virtual) $Z$ decay depends on the
mass splitting between $A$ and $H$ and is relatively small,
$M_{\mu\mu} \le M_{Z}$. 
We apply pre-selection cuts on the invariant mass  and the
longitudinal boost of the lepton pair to suppress the dominant
background process $e^+ e^- \to \mu^+ \mu^- (\gamma)$, see
Fig.~\ref{fig:presel}. 
Distributions of selected kinematic variables describing the
leptonic final state in $AH$ analysis, after the pre-selection cuts,
are presented in Fig.~\ref{fig:dist}.
Presented in Fig.~\ref{fig:bdt}\,(left) is the lepton pair invariant mass
distribution after pre-selection cuts and additional selection based
on lepton pair energy, transverse momentum, production angle (polar
angle of the $Z$ boson) and the difference of the lepton azimuthal angles.
Already with this simplest, cut-based approach, the IDM signal would
result in the visible excess in the invariant mass distribution for
the number of benchmark scenarios.
For the final selection of signal-like events, a multivariate analysis
is performed using a Boosted Decision Tree (BDT) classifier
\cite{Hocker:2007ht} with 8 input variables~\cite{Kalinowski:2018kdn}.
The standard approach in this type of analysis is to train BDT to
separate the considered signal scenario from the background events.
However, this approach, also used in our previous
study~\cite{Kalinowski:2018kdn}, is only valid if we do have some
initial estimates on the model parameters, scalar masses in particular.
For the results presented here we modified our approach and we train
BDTs using all accessible (at given energy) benchmark scenarios from
given category (separately for virtual and real $Z$ in the final
state) but for the one we look for. 
This procedure, which we consider a more general
(``scenario-independent'') approach, results in the
expected significances of observation lower by up to 20\% compared to
the ``educated-selection'' results.

\begin{figure}[tb]
\centerline{\includegraphics[width=0.8\textwidth]{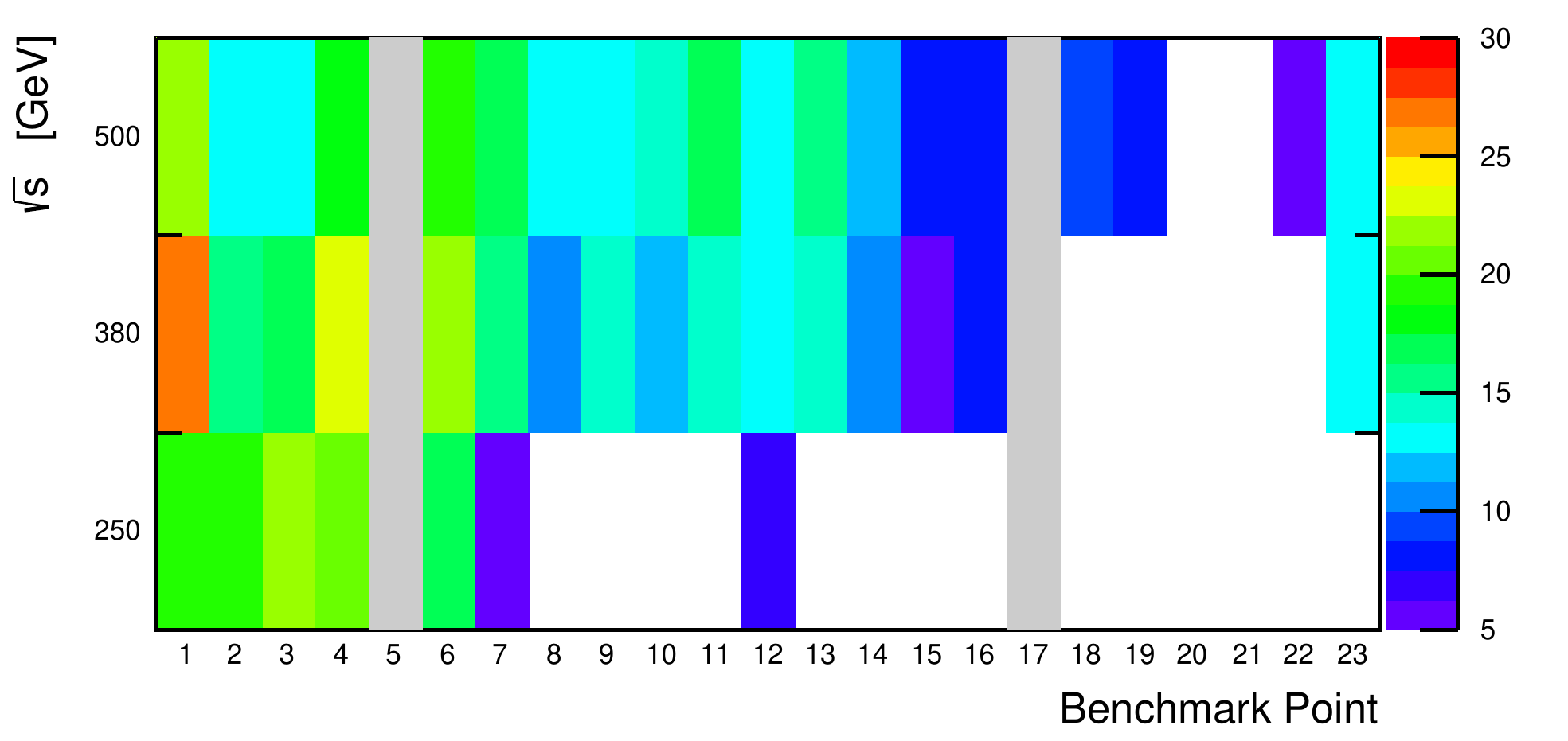}}
\caption{Significance of the deviations from the Standard Model
  predictions, expected for 1\,ab$^{-1}$
  of data collected at centre-of-mass energy of 250\,GeV, 380\,GeV and
  500\,GeV, for events with two muons in the final state, for all considered
  low mass benchmark scenarios. Only significance above 5$\sigma$ is shown.
}\label{fig:ahsig}
\end{figure}

Response distributions of the BDT classifier used for the
selection of $AH$ production events for the benchmark scenario BP1 at
250\,GeV are presented in Fig.~\ref{fig:bdt}\,(right).
Expected significance of the deviations from the Standard Model 
predictions, assuming  1\,ab$^{-1}$ of data collected at
centre-of-mass energy of 250\,GeV, 380\,GeV and 500\,GeV, are shown in Figs.~\ref{fig:ahsig}.
Only scenarios resulting in significances above 5$\sigma$ are
shown. 

\begin{figure}[tb]
\includegraphics[width=0.49\textwidth]{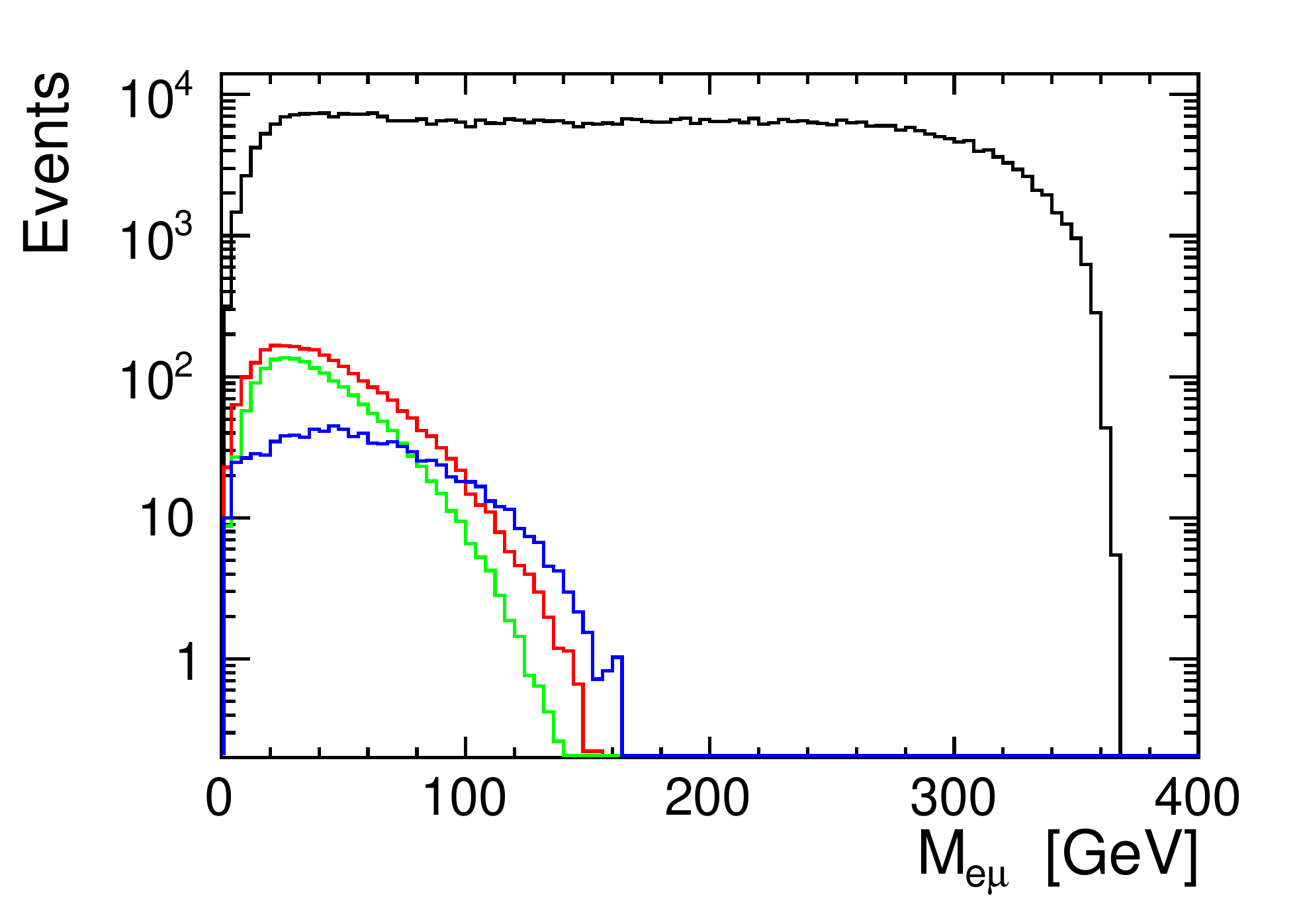}
\includegraphics[width=0.49\textwidth]{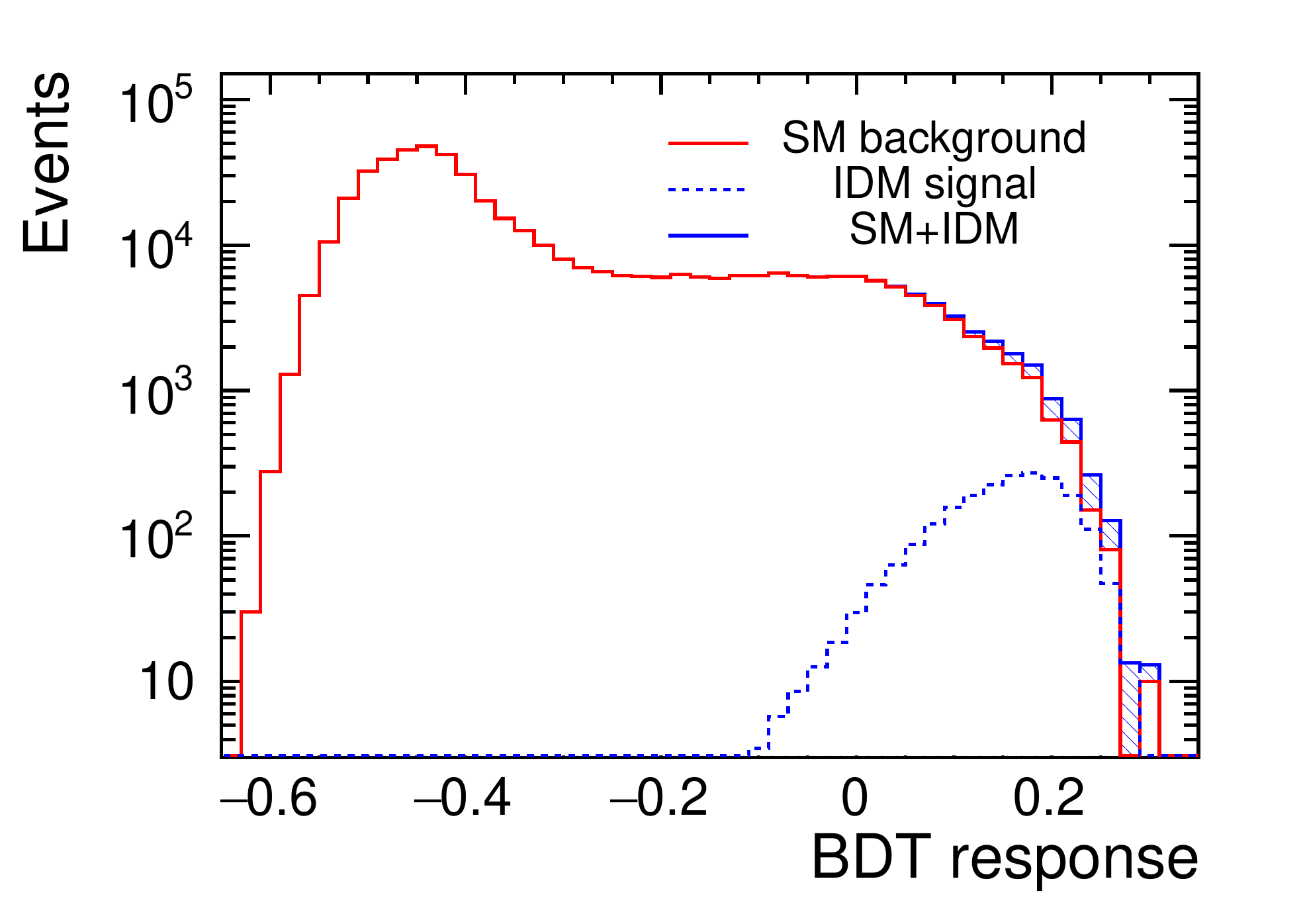}
\caption{
  Left: distribution of the lepton pair invariant mass, M$_{e\mu}$,
  for BP1 (red histogram), BP3 (green) and BP6 (blue) signal
  scenarios, compared with the expected Standard Model background (black
  histogram).
  Right: response distributions of the BDT classifiers used for the
  selection of $H^+H^-$ production events, for BP1.
  Samples are normalised to 1\,ab$^{-1}$ collected at 380\,GeV.
}\label{fig:bdt2}
\end{figure}

The selection of $H^+H^-$ production events is more challenging as the
two leptons in the final state no longer originate from a single (on-
or off-shell) intermediate state. 
No pre-selection cuts are applied (except for the detector acceptance
cuts on the generator level), but we focus on electron-muon pairs in
the final state, avoiding large SM background from the direct lepton pair production. 
In Fig.~\ref{fig:bdt2} (left) the distribution of the lepton pair
invariant mass, M$_{e\mu}$, for three benchmark scenarios (BP1, BP3
and BP6) is compared with Standard Model expectations for centre-of-mass energy of 380\,GeV.
The expected background cross section for the considered final state
is over two orders of magnitude higher than for the considered
benchmark points.
However, kinematic distributions are very different, as two massive
scalars are produced in the signal case, reducing the kinematic space
available for lepton pair production,
allowing for efficient selection of signal-enhanced sample of events
using the multivariate analysis. The same procedure and the same set
of input variables is used as for the $AH$ analysis.

\begin{figure}[tb]
\centerline{\includegraphics[width=0.8\textwidth]{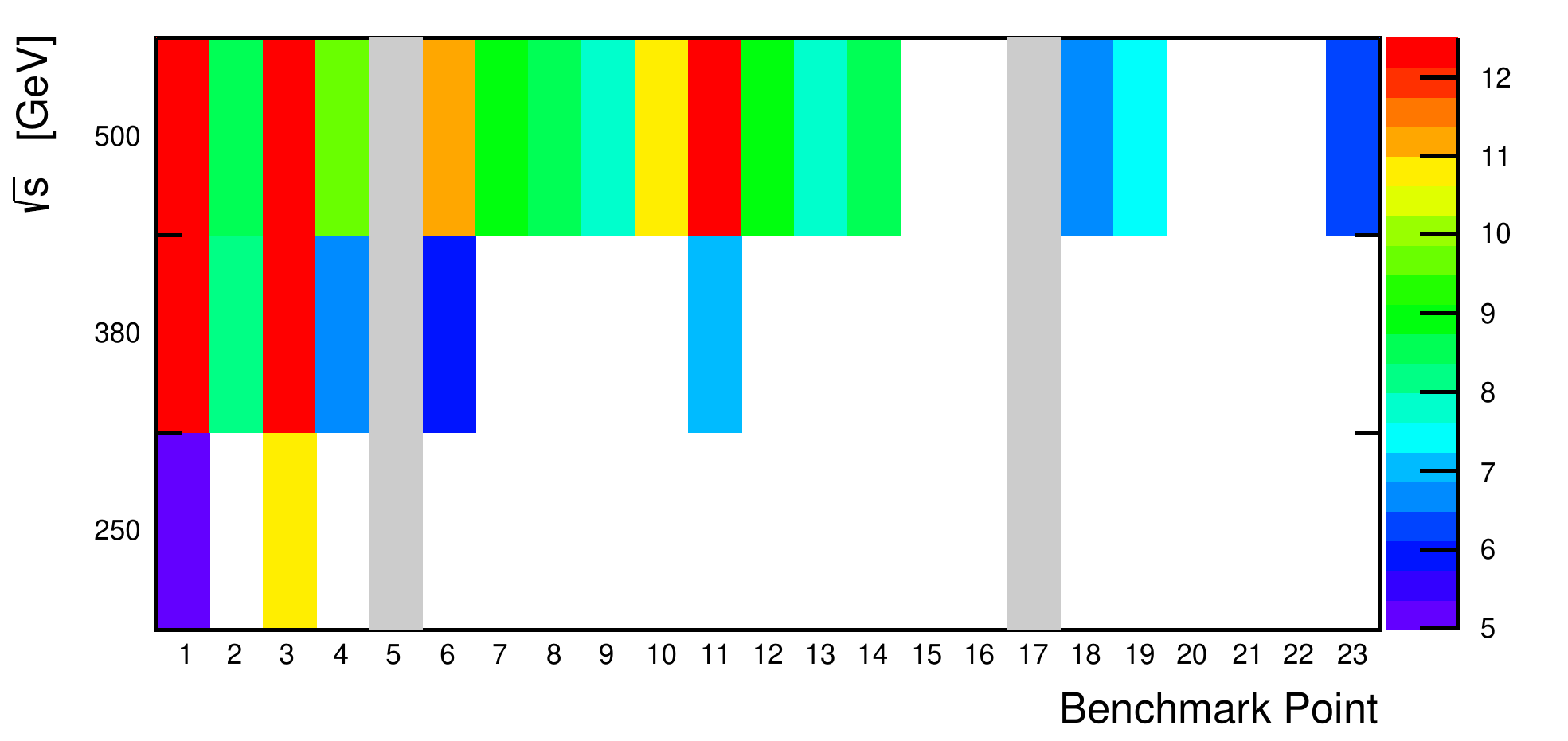}}
\caption{Significance of the deviations from the Standard Model
  predictions, expected for 1\,ab$^{-1}$
  of data collected at centre-of-mass energy of 250\,GeV, 380\,GeV and
  500\,GeV, for events with  an electron  and a muon in the final state, for all considered
  low mass benchmark scenarios. Only significance above 5$\sigma$ is shown.
}\label{fig:hphmsig}
\end{figure}

Response distributions of the BDT classifier used for the
selection of $H^+H^-$ production events for the benchmark scenario BP1
at 380\,GeV are presented in Fig.~\ref{fig:bdt2}\,(right).
In Fig.~\ref{fig:hphmsig} we show the expected significance of the
deviations from the Standard Model predictions for  1\,ab$^{-1}$ of
data collected at  250\,GeV, 380\,GeV and 500\,GeV,  
for scenarios resulting in the significances above 5$\sigma$. 

\begin{figure}[tb]
\hspace{0.03\textwidth}
  \includegraphics[width=0.47\textwidth]{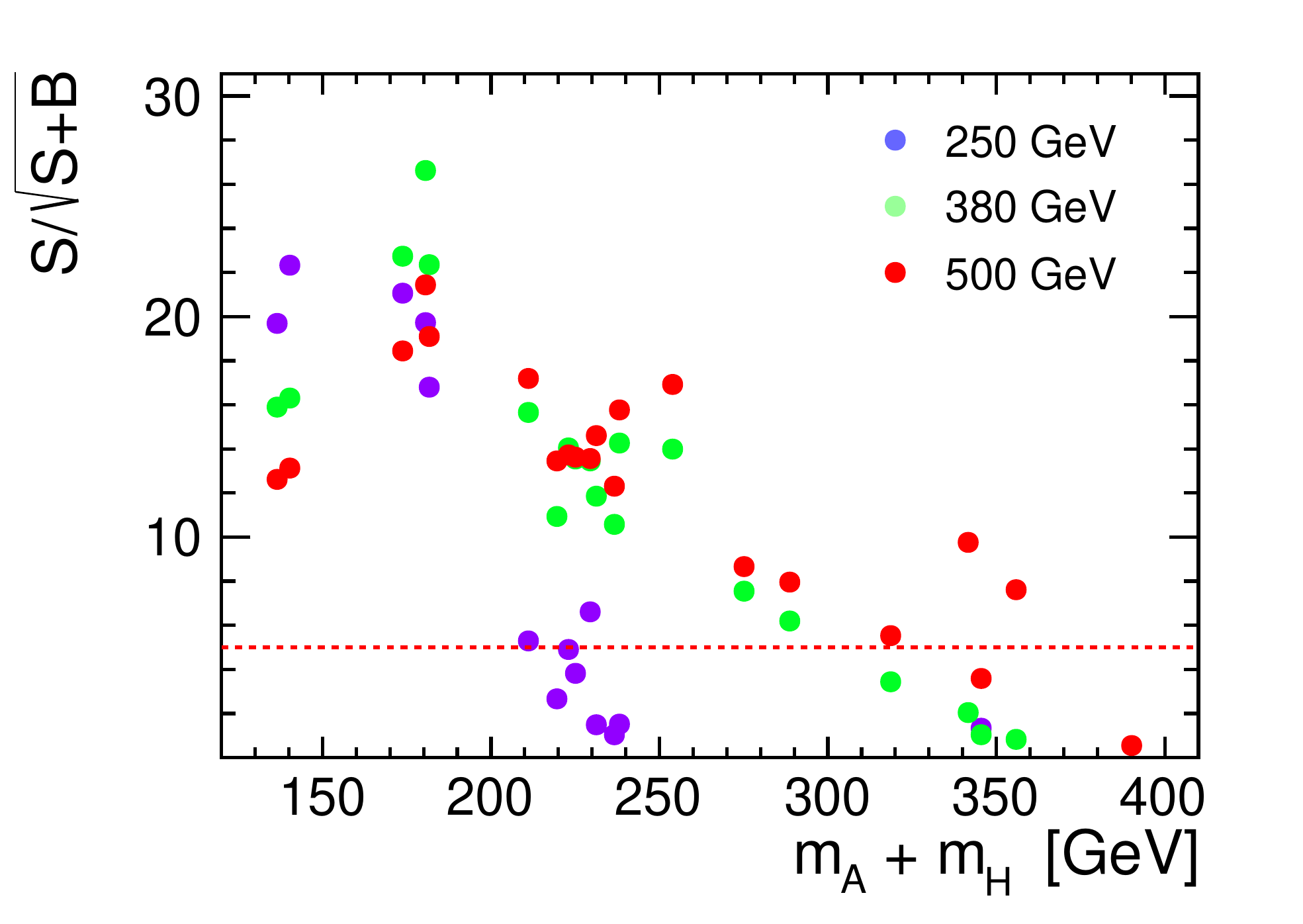}
  \includegraphics[width=0.47\textwidth]{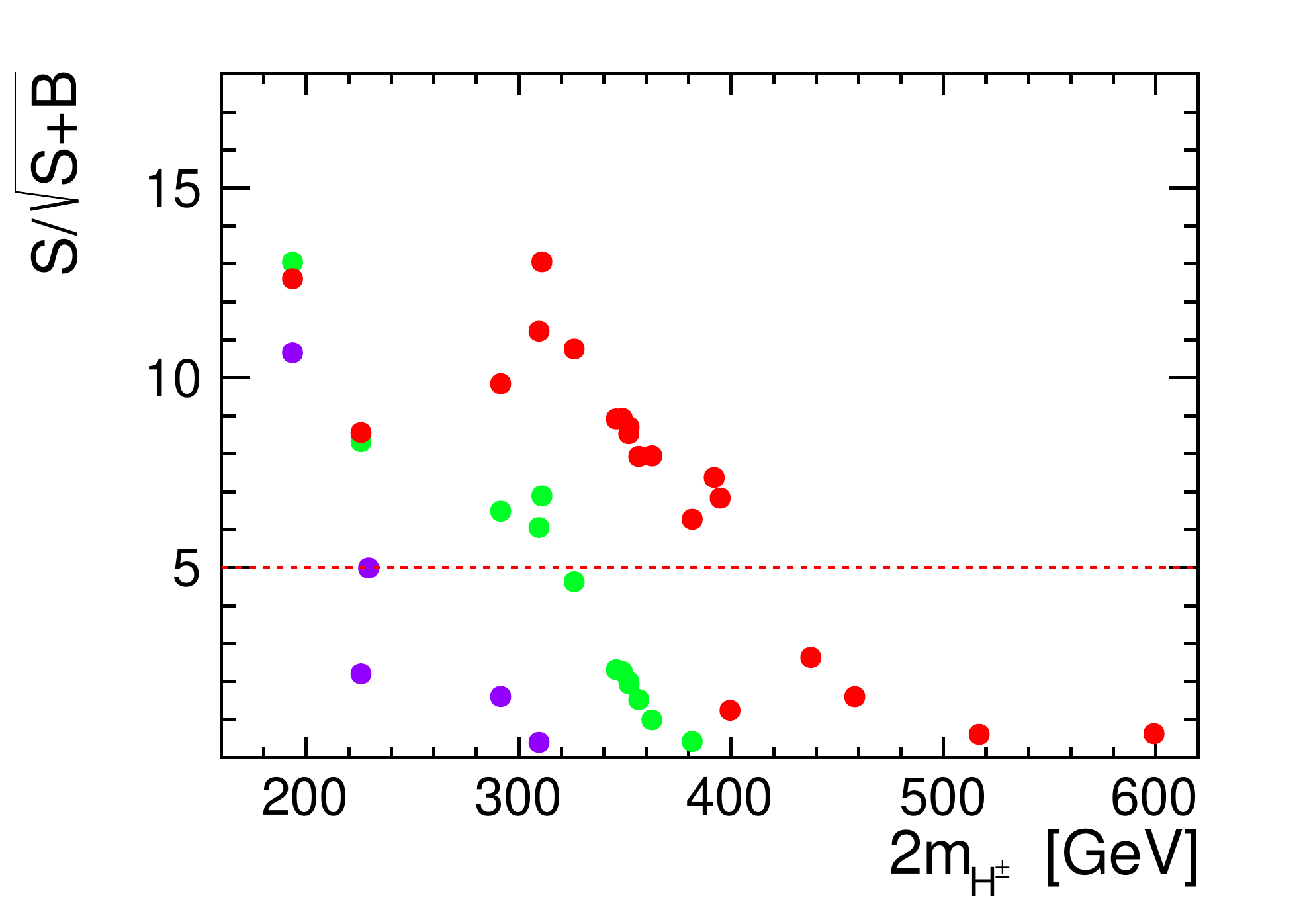}
\caption{Significance of the deviations from the Standard Model
  predictions expected for 1\,ab$^{-1}$ of data collected at
  centre-of-mass energy of 250\,GeV, 380\,GeV and   500\,GeV, for:
  (left) events with two muons in the final state ($\mu^+\mu^-$)
  as a function of the sum of neutral inert scalar masses and
  (right) events with an electron and a muon in the final state
  ($e^+\mu^-$ or $e^-\mu^+$) as a function of twice the charged scalar
  mass.  
}\label{fig:lesig}
\end{figure}

We found that for scenarios accessible at a certain energy, up to
500\,GeV, high significance can be expected for leptonic signature at
future $e^+e^-$ colliders. 
The significance is mainly related to the inert scalar production cross
sections.
We display the dependence of the expected significance on the inert
scalar masses, for events with two muons and for events with and
electron-muon pair in the final state, in Fig.~\ref{fig:lesig}.
With  1\,ab$^{-1}$ of integrated luminosity collected at 250\,GeV,
380\,GeV and   500\,GeV, the expected discovery reach of $e^+e^-$
colliders extends up to neutral scalar mass sum of 220\,GeV, 300\,GeV
and  330\,GeV, respectively, and for charged scalar pair-production up
to charged scalar masses of 110\,GeV, 160\,GeV and  200\,GeV.
\begin{figure}[tb]
 \hspace{0.03\textwidth}
  \includegraphics[width=0.47\textwidth]{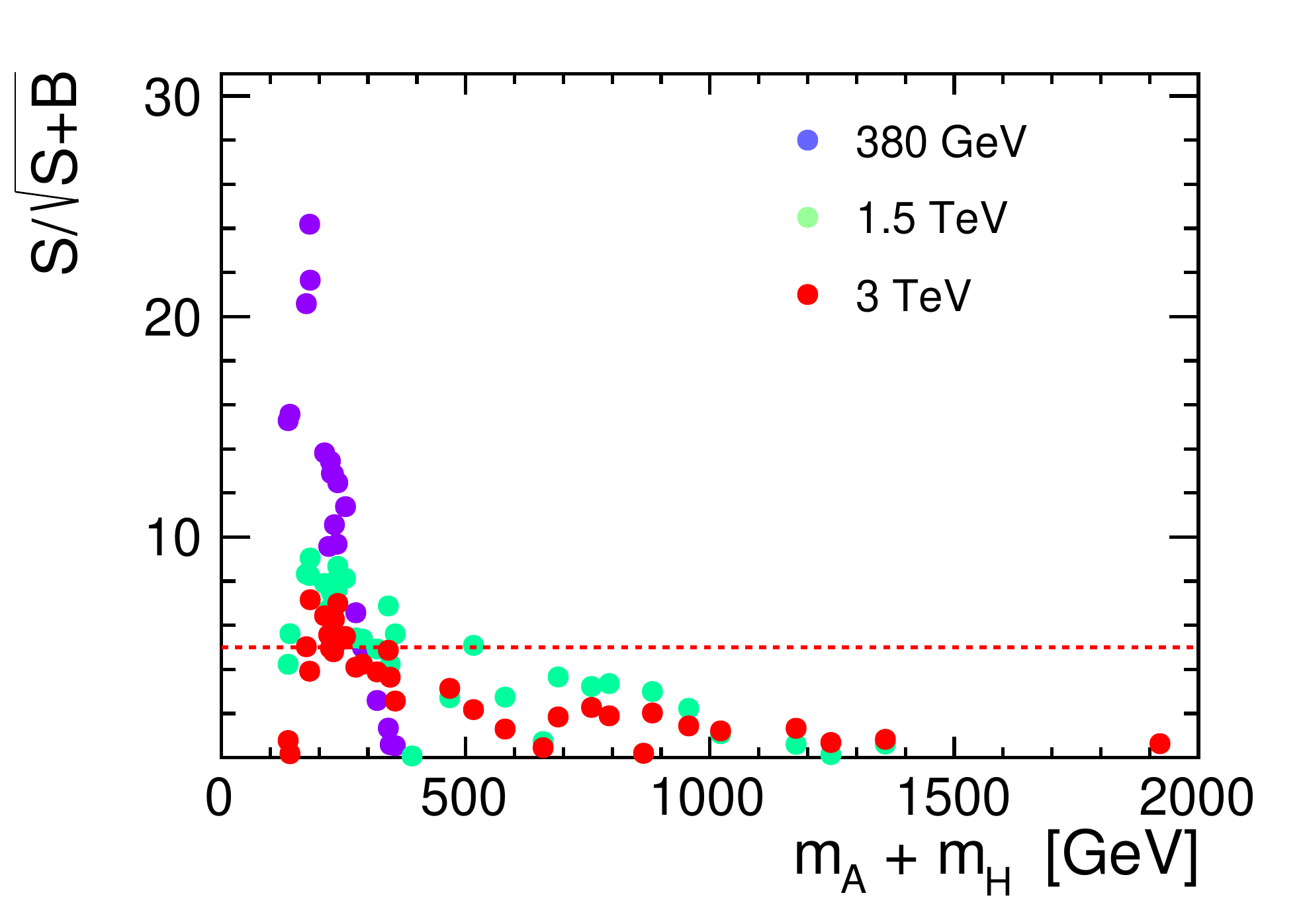}
  \includegraphics[width=0.47\textwidth]{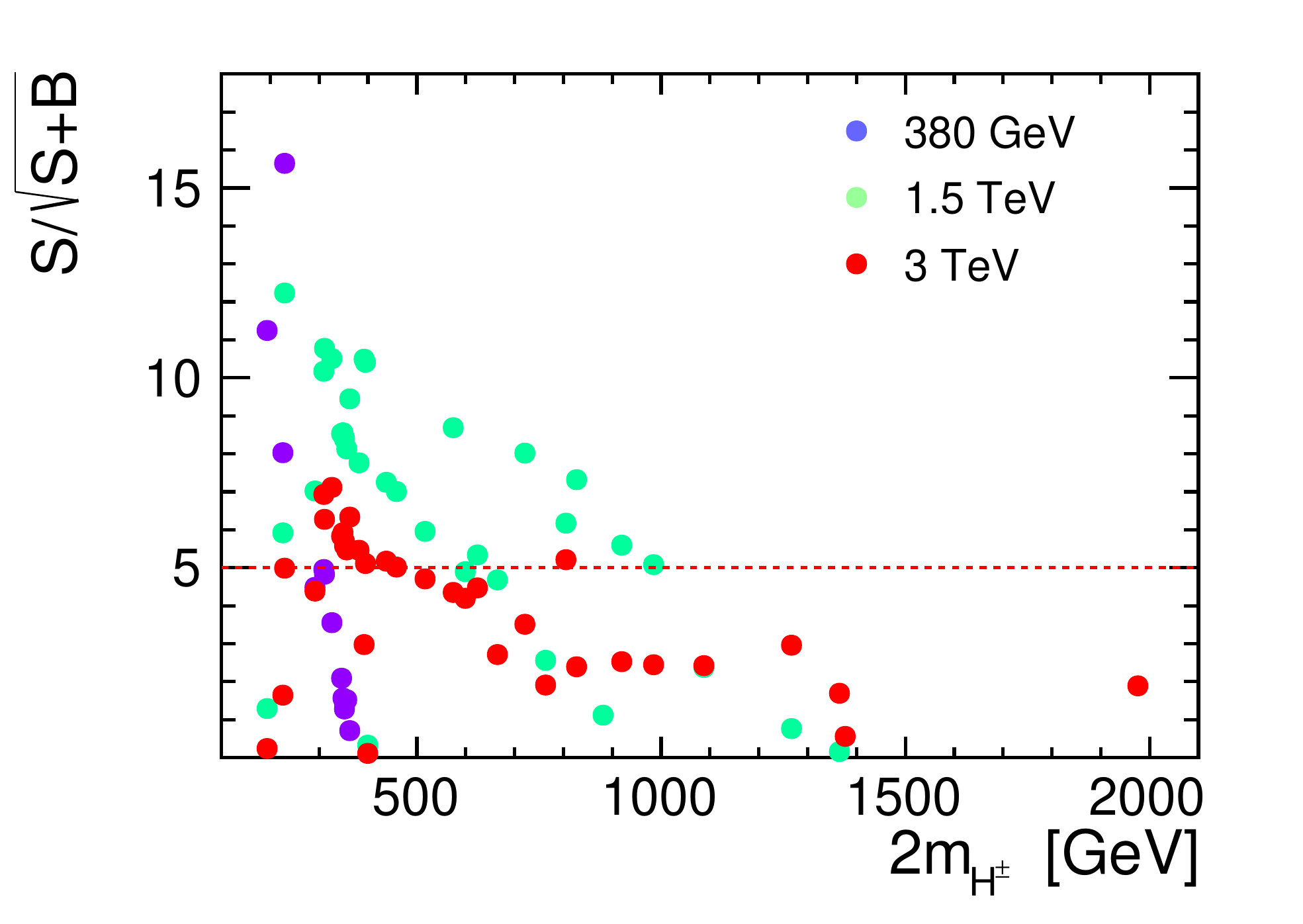}
  \caption{As in Fig.~\ref{fig:lesig} but for expected CLIC running
    scenario: 1\,ab$^{-1}$ of data collected at 380\,GeV,
    2.5\,ab$^{-1}$ at 1.5\,TeV and 5\,ab$^{-1}$ at 3\,TeV.
  }\label{fig:hesig}
\end{figure}

For collision energies much above the threshold, the inert scalar
pair-production cross section decreases fast with the collision energy
(see Fig.~\ref{fig:cros}).
For CLIC running at 1.5\,TeV, only a moderate increase in discovery
reach is expected for the leptonic channel, even with 2.5\,ab$^{-1}$
of data, see Fig.~\ref{fig:hesig}.
The neutral scalar pair-production can be discovered in the leptonic
channel for $m_A + m_H < 450$\,GeV
and the charged scalar production for $m_{H^\pm} < 500$\,GeV.
Marginal improvement is expected when running at 3 TeV.

\begin{figure}[tb]
 \hspace{0.03\textwidth}
  \includegraphics[width=0.47\textwidth]{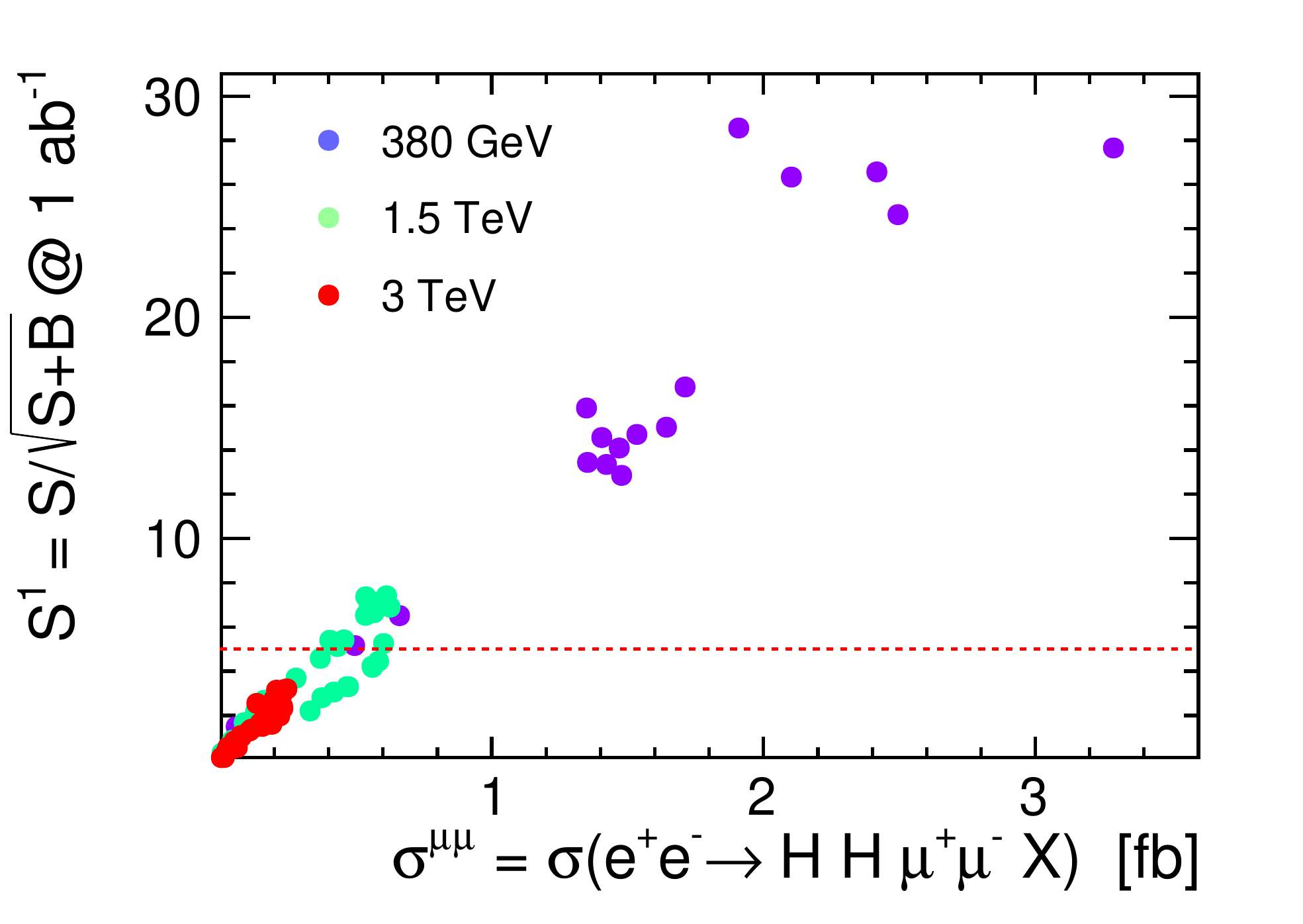}
  \includegraphics[width=0.47\textwidth]{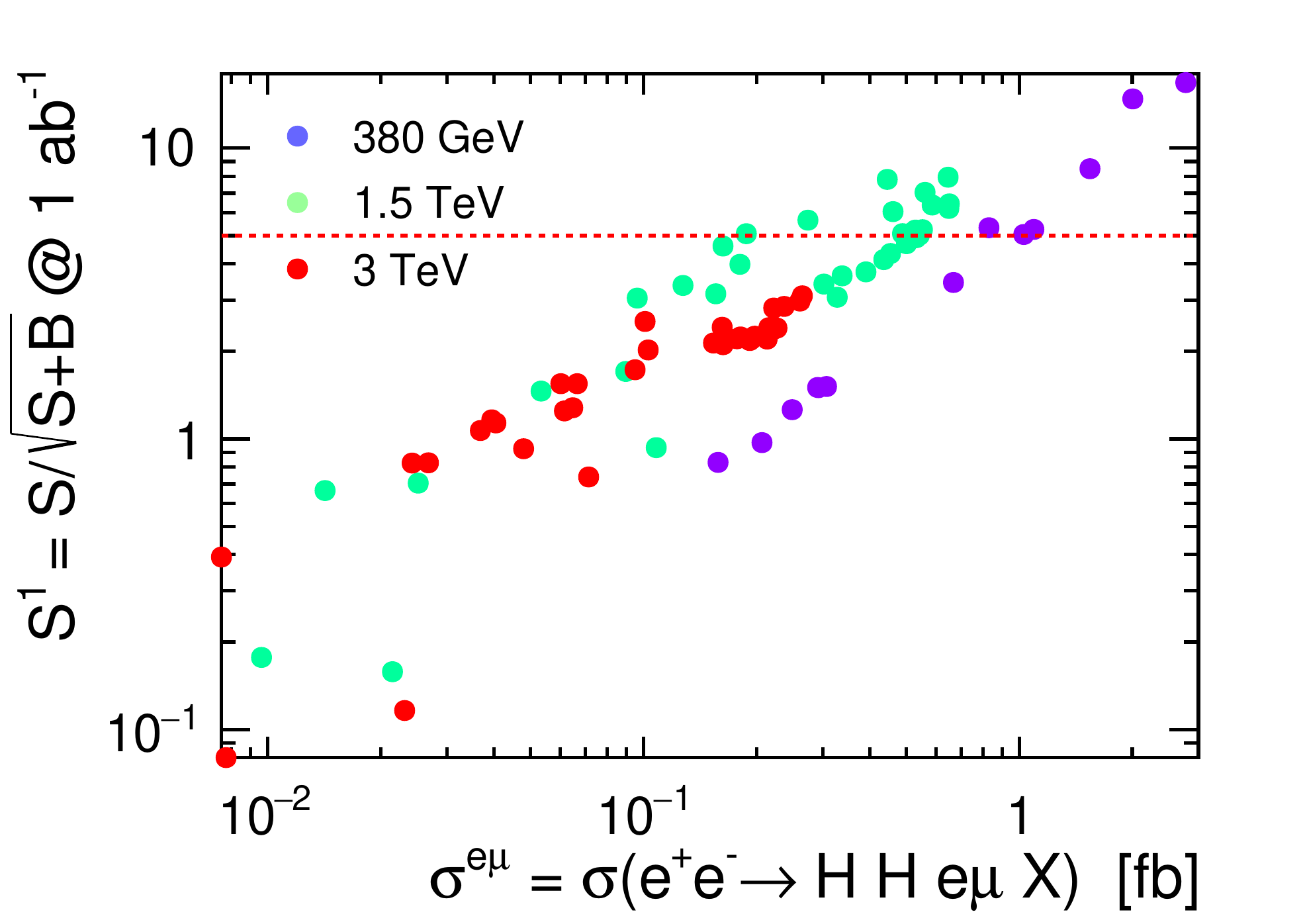}
  \caption{Significance of the deviations from the Standard Model
  predictions expected at different CLIC running stages, assuming
  the same integrated luminosity of 1 ab$^{-1}$, as a function of
  the signal cross section in the considered channel, for:
  (left) events with two muons in the final state ($\mu^+\mu^-$)
  and
  (right) events with an electron and a muon in the final state
  ($e^+\mu^-$ or $e^-\mu^+$).
  }\label{fig:hesig2}
\end{figure}

The significance is mainly driven by the signal production cross section and
is approximately proportional to the square-root of the integrated luminosity.
Shown in Fig.~\ref{fig:hesig2} are the significance results scaled to
the integrated luminosity of 1\,ab$^{-1}$, presented as a function of
the signal production cross section.
For the $AH$ channel, which leads to $\mu^+\mu^-$ final state, a
universal linear dependence on the signal cross section is observed
which does not seem to depend on the running energy. Significant
(above $5\sigma$) observation is possible for cross sections
roughly larger than 0.5 fb.
For the $H^+H^-$ channel, with $e^\pm \mu^\mp$ final state, low energy
running seem to give better sensitivity to signal scenarios for the
same cross section. 

\section{Semi-leptonic channel}

\begin{figure}[tb]
\centerline{\includegraphics[width=0.4\textwidth]{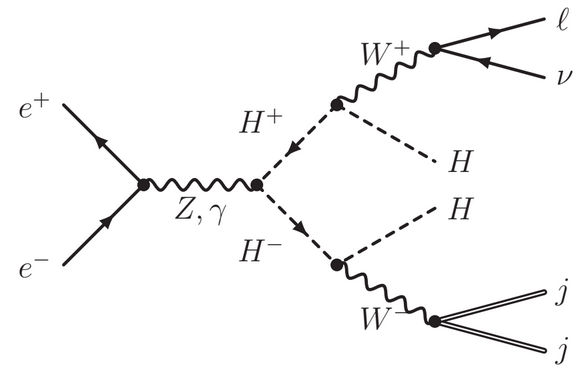}}
\caption{Signal Feynman diagram for the charged scalar pair-production in semi-leptonic decay channel:
$e^+e^- \to H^+ H^- \to H H j j l \nu$.
}\label{fig:diag2}
\end{figure}

For charged scalar pair-production, significant improvement of the discovery reach
for scenarios with high scalar masses can be achieved using the
semi-leptonic final state, see Fig.~\ref{fig:diag2}.
As the signal cross section increases by an order of magnitude and 
a similar scaling is expected for the background processes (dominated
by the $W^+W^-$ production), the significance of the
observation in the semi-leptonic channel should increase by a factor
of about 3.
Additional improvement is possible due to kinematic constraints which
can be imposed on the hadronic final state (corresponding to one of
the produced $W$ bosons).
However, detector response has to be taken into account in more
details.

Results presented in the following are based on the signal and
background event samples were generated with \whizard
2.7.0~\cite{Kilian:2007gr,Ohl:2000hq}, taking into account the beam
energy profile expected for CLIC running at 1.5\,TeV and 3\,TeV.
We assume running with -80\% electron beam polarisation and the
corresponding integrated luminosity of 2\,ab$^{-1}$ and 4\,ab$^{-1}$
respectively.
For realistic simulation of the CLIC detector response fast simulation
framework \delphes~\cite{deFavereau:2013fsa} was used, with control
cards prepared for the new detector model CLICdet~\cite{Leogrande:2019qbe}. 

\begin{figure}[tb]
\includegraphics[width=0.49\textwidth]{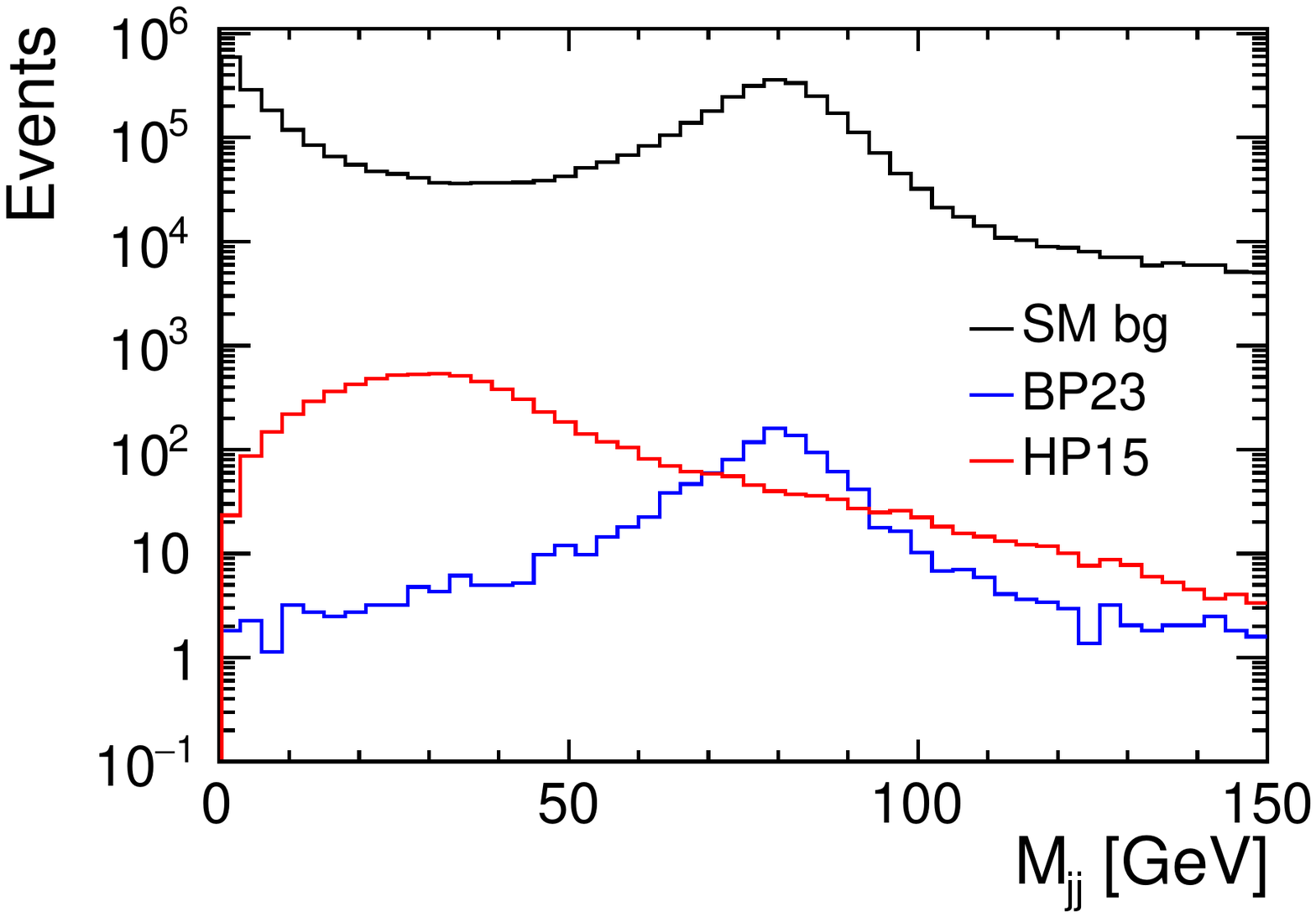}
\includegraphics[width=0.49\textwidth]{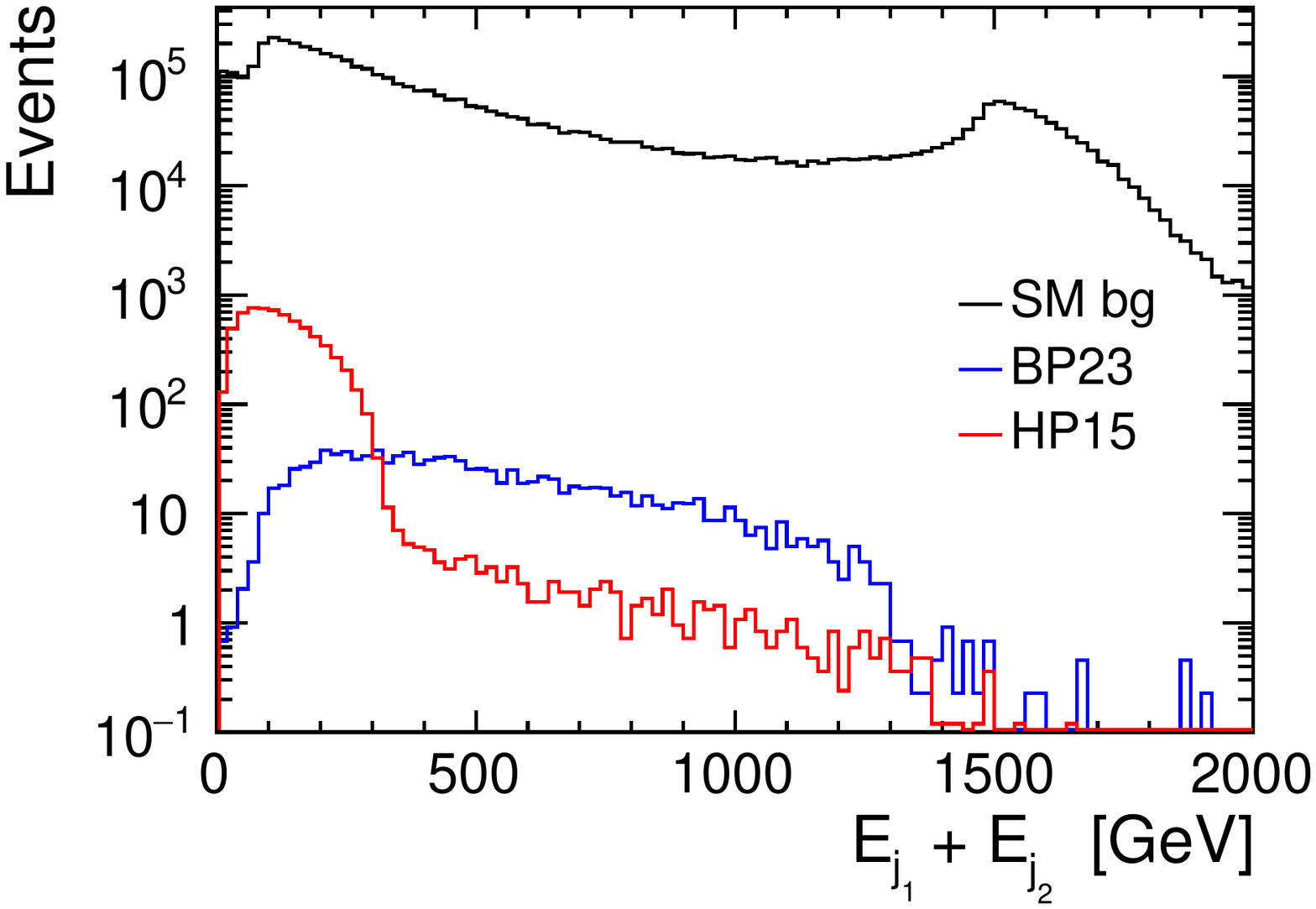}
\caption{
  Distributions of the kinematic variables describing the
  semi-leptonic final state in $H^+ H^-$ analysis: jet pair invariant mass,
  $M_{jj}$, and the sum of jet energies, $E_{j_1} + E_{j_2}$.
Expected distributions for benchmark scenarios BP23 (blue
histogram) and HP15 (red) are compared with expected
background (black histogram) simulated for 4\,ab$^{-1}$ of data
collected at 3\,TeV width -80\% electron beam polarisation.
}\label{fig:slepdist}
\end{figure}

Selected for the analysis are events with exactly one isolated lepton
(electron or muon) and two exclusive jets reconstructed with the VLC
algorithm\footnote{The VLC algorithm is run with parameter $R=1$ at 1.5\,TeV
  and $R=1.2$ at 3\,TeV, and with $\beta=\gamma=1$.}
\cite{Boronat:2016tgd}.
Also rejected are events with an isolated photon with energy above 10 GeV
or with the energy sum of the energy-flow objects outside the two
reconstructed jets higher than 20 GeV.
In Fig.~\ref{fig:slepdist}, distributions of the jet pair invariant mass,
  $M_{jj}$, and the sum of jet energies, $E_{j_1} + E_{j_2}$, for the
two signal scenarios, are compared with the expected SM background for
CLIC running at 3\,TeV.

The analysis procedure is similar to the one used for the leptonic
channel.
Huge background coming mainly from $W^+W^-$ and $ZZ$
pair-production is first suppressed by the pre-selection cuts based on
lepton and jet kinematics.
Then a multivariate analysis is performed using the BDT classifier
with 11 input variables:
total energy in an event, missing transverse momentum, missing
(recoil) mass; energy, transverse momentum and scattering angle of the
isolated lepton; energy, invariant mass and emission angle of the jet
pair; reconstructed angles of the hadronic $W$ decay.
As before, the BDT is trained separately for scenarios with virtual
$W^\pm$ production (when the difference of $H^\pm$ and $H$ masses is
smaller than the mass of $W^\pm$) and with real $W^\pm$ production
(larger mass differences).

\begin{figure}[tb]
\centerline{\includegraphics[width=0.55\textwidth]{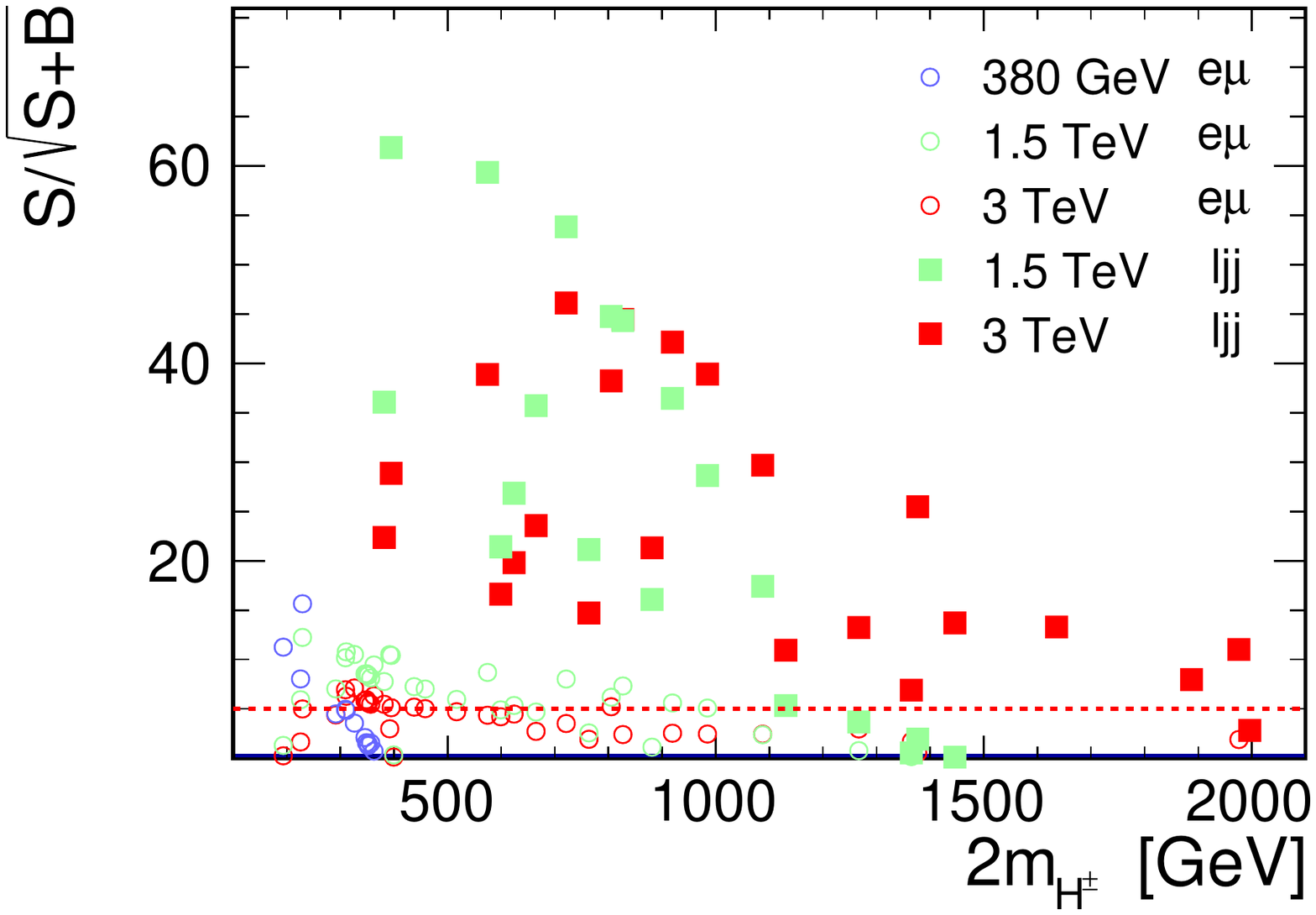}}
\caption{Significance of the deviations from the Standard Model predictions in the
  leptonic channel (open circels) and the semi-leptonic channel (filled squares)
  as a function of twice the charged scalar mass for expected CLIC running
   scenario: 1\,ab$^{-1}$ of data collected at 380\,GeV,
   2.5\,ab$^{-1}$ at 1.5\,TeV and 5\,ab$^{-1}$ at 3\,TeV.
}\label{fig:sigslep}
\end{figure}

Shown in Fig.~\ref{fig:sigslep} is the significance for observing
deviations from the Standard Model predictions.
Results based on the semi-leptonic channel analysis for CLIC running
at 1.5\,TeV and 3\,TeV are compared with the leptonic channel
sensitivity presented in Sec.~\ref{sec:leptonic}.
Huge increase of the signal significance is observed, up to a factor
of 6, and the discovery reach for charged scalar pair-production is
extended up to $m_{H^\pm} \sim$ 1\,TeV. 

\section{Conclusions}

The Inert Doublet Model is one of the simplest SM extensions
providing natural candidate for dark matter.
Light IDM scenarios, with scalar masses in $\mathcal{O}$(100\,GeV)
range are still not excluded by the current experimental and
theoretical constraints. 
Low mass IDM scenarios can be observed with high significance in the
di-lepton channels already at a $e^+e^-$ collider with 250\,GeV
center-of-mass energy.
Discovery reach increases for higher $\sqrt{s}$ and significant
improvement in the discovery reach is observed when considering the
semi-leptonic final state. 
Full simulation study of the charge scalar pair-production in the
semi-leptonic decay channel is ongoing.

\subsection*{Acknowledgements}

This contribution was supported by the National Science Centre, Poland, the
OPUS project under contract UMO-2017/25/B/ST2/00496 (2018-2021) and
the HARMONIA project under contract UMO-2015/18/M/ST2/00518
(2016-2019), and by the European Regional Development Fund - the
Competitiveness and Cohesion Operational Programme (KK.01.1.1.06 - RBI
TWIN SIN).

\end{document}